\documentclass[11pt,a4paper]{article}

\usepackage{authblk}
\usepackage[sort&compress,numbers]{natbib}
\usepackage{amsmath}
\usepackage{caption}
\usepackage{graphicx}
\usepackage{booktabs}
\usepackage[hidelinks]{hyperref}
\usepackage{cleveref}
\usepackage[left=1.5cm,right=1.5cm,top=1.5cm,bottom=2.5cm]{geometry}

\title{Climate uncertainty impacts on optimal mitigation pathways and social cost of carbon}

\author[1,2]{Christopher J. Smith}
\author[2,3]{Alaa Al Khourdajie} 
\author[4,5]{Pu Yang}
\author[6]{Doris Folini}

\affil[1]{Priestley International Centre for Climate, University of Leeds, Woodhouse Lane, Leeds, LS2 9JT, United Kingdom}
\affil[2]{Energy, Climate and Environment Program, International Institute for Applied Systems Analysis, Schlo\ss platz 1, 2361 Laxenburg, Austria}
\affil[3]{Centre for Environmental Policy, Imperial College, South Kensington Campus, London, SW7 2AZ, United Kingdom}
\affil[4]{Energy and Power Group, University of Oxford, OX5 1PF, United Kingdom}
\affil[5]{Bartlett School of Sustainable Construction, University College London, Gower Street, London, WC1E 6BT, United Kingdom}
\affil[6]{Institute for Climate and Atmospheric Science, ETH Z\"{u}rich, Universit\"{a}tstrasse 16, 8092 Z\"{u}rich, Switzerland}

\begin{document}
\maketitle


\begin{abstract} 

Emissions pathways used in climate policy analysis are often derived from integrated assessment models. However, such emissions pathways do not typically include climate feedbacks on socioeconomic systems and by extension do not consider climate uncertainty in their construction. We use a well-known cost-benefit integrated assessment model, the Dynamic Integrated Climate-Economy (DICE) model, with its climate component replaced by the Finite-amplitude Impulse Response (FaIR) model (v2.1). The climate uncertainty in FaIR is sampled with an ensemble that is consistent with historically observed climate and Intergovernmental Panel on Climate Change (IPCC) assessed ranges of key climate variables such as equilibrium climate sensitivity (ECS). By varying discounting assumptions, three scenarios are produced: a pathway similar to the ``optimal welfare'' scenario of DICE that has similar warming outcomes to current policies, and pathways that limit warming to ``well-below'' 2°C and 1.5°C with a short-term overshoot, aiming to meet Paris Agreement long-term temperature goals. Climate uncertainty alone is responsible for a factor of five variation (5--95\% range) in the social cost of carbon (SCC) in the 1.5°C overshoot scenario, with the spread in SCC increasing in relative terms with increasing stringency of climate target. CO$_2$ emissions trajectories resulting from the optimal level of emissions abatement in all pathways are also sensitive to climate uncertainty, with 2050 emissions ranging from \textminus12 to +14 GtCO$_2$ yr$^{-1}$ in the 1.5°C scenario. ECS and the strength of present-day aerosol effective radiative forcing are strong determinants of SCC and mid-century CO$_2$ emissions. This shows that narrowing climate uncertainty leads to more refined estimates for the social cost of carbon and provides more certainty about the optimal rate of emissions abatement. Including climate and climate uncertainty in integrated assessment model derived emissions scenarios would address a key missing feedback in scenario construction.
\end{abstract}

\section{Introduction}

Integrated Assessment Models (IAMs) can be categorized into two broad types: process-based (PB-IAMs) and cost-benefit (CB-IAMs) \cite{Weyant2017}. PB-IAMs model the energy system, technology, economy, agricultural productivity and land use across a number of world regions, are used to construct possible future emissions scenarios, and have extensive policy reach \cite{vanBeek2020}, partly as a consequence of their ubiquity across IPCC reports \cite{Riahi2022}. PB-IAMs produced the Shared Socioeconomic Pathways (SSPs) used to drive Earth System model projections of future climate \cite{ONeill2016}, providing a large base of model evidence to the Intergovernmental Panel on Climate Change (IPCC) Working Group 1 (WG1) report. Analysis of future potential technological and social developments in a large number of PB-IAMs are assessed in IPCC Working Group 3 (WG3) \cite{Riahi2022}.

CB-IAMs are simpler and often used to model climate change effects on the global economy at a macro level. One area in which CB-IAMs have had extensive policy reach is in determining the social cost of carbon (SCC), describing the marginal time-discounted climate damages suffered by society for each additional ton of CO$_2$ emitted \cite{Weyant2017}. CB-IAMs perform a cost-benefit analysis that balances the foregone present-day economic consumption (which under the current global energy mix, is CO$_2$-intensive) that is instead invested in emissions abatement technologies, with benefits future avoided climate damages from warming. The SCC forms a central component of climate policy in several countries, most notably the United States \cite{Rennert2022}. In a hypothetical efficient market, the SCC could be used to set the optimal global carbon price or carbon taxation level. 

A CB-IAM requires a simple climate module as an integral part of the model in order to calculate global warming and hence climate damages. While their model dynamics are highly aggregated and parameterised, CB-IAMs tend to include a two-way coupling between emissions and climate.
  PB-IAMs may also include climate modules and may calculate climate damages \cite{Bosetti2006} allowing determination of SCC, and may also be run in an optimization framework in order to produce cost-optimal energy transition pathways \cite{Huppmann2019}, but at present typically do not consider climate change effects on technology availability, energy demand, agriculture, or the factors of productivity when used to construct community emissions scenarios \cite{vanVuuren2012,Calvin2018,vanRuijven2019}. This potentially excludes important feedbacks between climate and human decision-making in scenario design. 

Additionally, the relative simplicity of CB-IAMs means that an optimal solution (e.g. from an iterative optimization process) can be found relatively quickly. Therefore, uncertainty analysis can be undertaken by varying model parameters and re-running many times using variance-based sensitivity analyses or Monte Carlo sampling \cite{Anderson2014,Miftakhova2021}. The properties of economic-climate coupling and efficiency make CB-IAMs useful tools for exploring the impact of climate uncertainty on emissions scenarios and SCC.

It has recently been observed that climate module components of CB-IAMs are performing poorly with respect to full-complexity Earth System models and observations \cite{Dietz2021}. CB-IAM climate modules can be improved if model parameters are better calibrated \cite{Folini2022}, though key Earth System processes such as the carbon cycle feedback are often missing \cite{Woodard2019}. As climate damages (and therefore SCC) in CB-IAMs depend on global mean surface temperature, it is important to use an appropriate and well-calibrated simple climate model within a CB-IAM to prevent biased estimates of SCC \cite{Dietz2021}.

An additional consideration for SCC is that of uncertainty in climate. Several climate variables including ECS and the magnitude of present day aerosol forcing have large uncertainty bounds \cite{Forster2021} and varying the climate response in CB-IAMs can lead to differing estimates of the SCC \cite{Su2018,Wang2022,Rennert2022}. We extend this previous work by producing a systematic assessment of climate uncertainty using a calibrated probabilistic ensemble of the FaIR v2.1.0 simple climate model \cite{Leach2021} coupled to the DICE-2016R CB-IAM \cite{Nordhaus2017}, focusing on allowable CO$_2$ emissions under Paris Agreement consistent mitigation scenarios in addition to the effect of climate uncertainty on the present-day SCC.

\section{Methods}

\subsection{DICE integrated assessment model}
\label{sec:dice}
The starting point for this work is the DICE-2016R model of William Nordhaus \cite{Nordhaus2013,Nordhaus2017} with some additional updates and modifications (Supplementary Material sect. 1). We reduce the model timestep in DICE from 5 years to 3 years, and use 2023 as the first period (updated from 2015 in DICE-2016R). We run DICE to 2500 for a total of 160 periods (DICE-2016R runs to 2510 for a total of 100 periods). A 3-year time step allows for more responsive emissions reductions in the near term, without significantly adding to the computational burden. 

Gross world economic output $Y$ is determined with a Cobb-Douglas production function
\begin{equation}
Y(t) = A(t) K(t)^{\gamma} L(t)^{1-\gamma}
\label{eq:GDP}
\end{equation}
where $K$ is global capital stock, $L$ is global labour stock, $\gamma = 0.3$ is the output elasticity to capital and $A$ is total factor productivity. $t = 1 \ldots 160$ is the period. $L(t)$ is assumed to scale proportionally with global population.

The projections of world population from DICE-2016R are updated with the median projection of 10,000 scenarios from the Resources For the Future Socioeconomic Pathways (RFF-SPs) \cite{Rennert2022,Raftery2023}. Global capital stock $K(t)$ and total global product $Y(t)$ are updated to use 2019 figures from the International Monetary Fund (IMF) reported in 2017\$ and re-indexed to give $K=$ \$341tn and $Y=$ \$133tn for 2023 in 2020\$. Total factor productivity $A(t)$ in 2023 is calculated by rearrangement of \cref{eq:GDP} using the re-indexed 2019 estimates of $K(t)$ and $Y(t)$ from the IMF data and $L(t)$ from the RFF-SP timeseries.

CO$_2$ emissions from fossil fuel and industrial processes ($E_{\mathrm{FFI}}$) are given by
\begin{equation}
E_{\mathrm{FFI}}(t) = \sigma(t) Y(t) (1-\mu(t))
\label{eq:emissions}
\end{equation}
where $\sigma(t)$ is the emissions intensity of GDP [kg CO$_2$ \$$^{-1}$]. $\sigma(t)$ includes a baseline improvement in energy efficiency over time in the absence of any climate policy. 
We update $E_{\mathrm{FFI}}$ to be 36.6 Gt CO$_2$ yr$^{-1}$ in 2023, which is the estimate of 2022 fossil fuel emissions from the Global Carbon Project (GCP) \cite{Friedlingstein2022}.

$\mu(t)$ is the emissions abatement fraction. In DICE-2016R, net negative emissions ($\mu > 1$) are not allowed until 2160. We relax this assumption, allowing net zero CO$_2$ emissions ($\mu = 1$) in 2040 and net negative emissions thereafter. While the feasibility of achieving net zero CO$_2$ emissions in 2040 is debatable \cite{Anderson2016,Fuss2018,Gambhir2019}, many PB-IAM scenarios in the IPCC WG3 database have already reached net negative emissions by 2040 \cite{Riahi2022,Byers2022}. In order to construct sensible transition pathways, we impose an upper limit of $\mu(t) = 0.15t$ for $1 \leq t \leq 7$ and retain DICE-2016R's maximum allowable abatement of $\mu(t) = 1.2$ for $t \geq 8$. 
We use $\mu = 0.15$ in 2023 rather than DICE-2016R's $\mu = 0.03$ in 2015. 
A present-day emissions abatement level of 15\% can be justified on the basis that some limited emissions mitigation has occurred. Around 10\% of global primary energy supply is renewable \cite{IEA2022}, and a significant coal-to-gas shift has occurred over the last 30 years in the energy sector. 

Total CO$_2$ emissions are given by $E = E_{\mathrm{FFI}} + E_{\mathrm{AFOLU}}$. $E_{\mathrm{AFOLU}}$ is the CO$_2$ emissions from agriculture, forestry and other land use (AFOLU). 
DICE-2016R uses an exogenous pathway of AFOLU CO$_2$ emissions. We replace this with a regression-based relationship of $E_{\mathrm{AFOLU}}$ with $E_{\mathrm{FFI}}$ and $t$ that is derived from 1202 PB-IAM scenarios from the IPCC WG3 database (Supplementary Material sect. 1.3).

\subsection{The calibrated FaIR v2.1 climate model}

FaIR is described in refs. \cite{Millar2017fair,Smith2018,Leach2021}. Unlike the DICE-2016R climate module, FaIR includes carbon cycle feedbacks simulating the declining efficiency of land and ocean carbon sinks (increasing airborne fraction) with increasing emissions of CO$_2$. A recent update, DICE-2023R, incorporates FaIR inside its climate module and therefore does include carbon cycle feedbacks. Comparisons with DICE-2016R and DICE-2023R are shown in Supplementary fig. 6. The version of FaIR v2.1.0 used inside DICE is a reduced version that includes just the carbon cycle and temperature response to forcing (Supplementary Material sect. 2).

\begin{figure*}[tbhp]
\centering
\includegraphics[width=17.8cm,height=8.9cm]{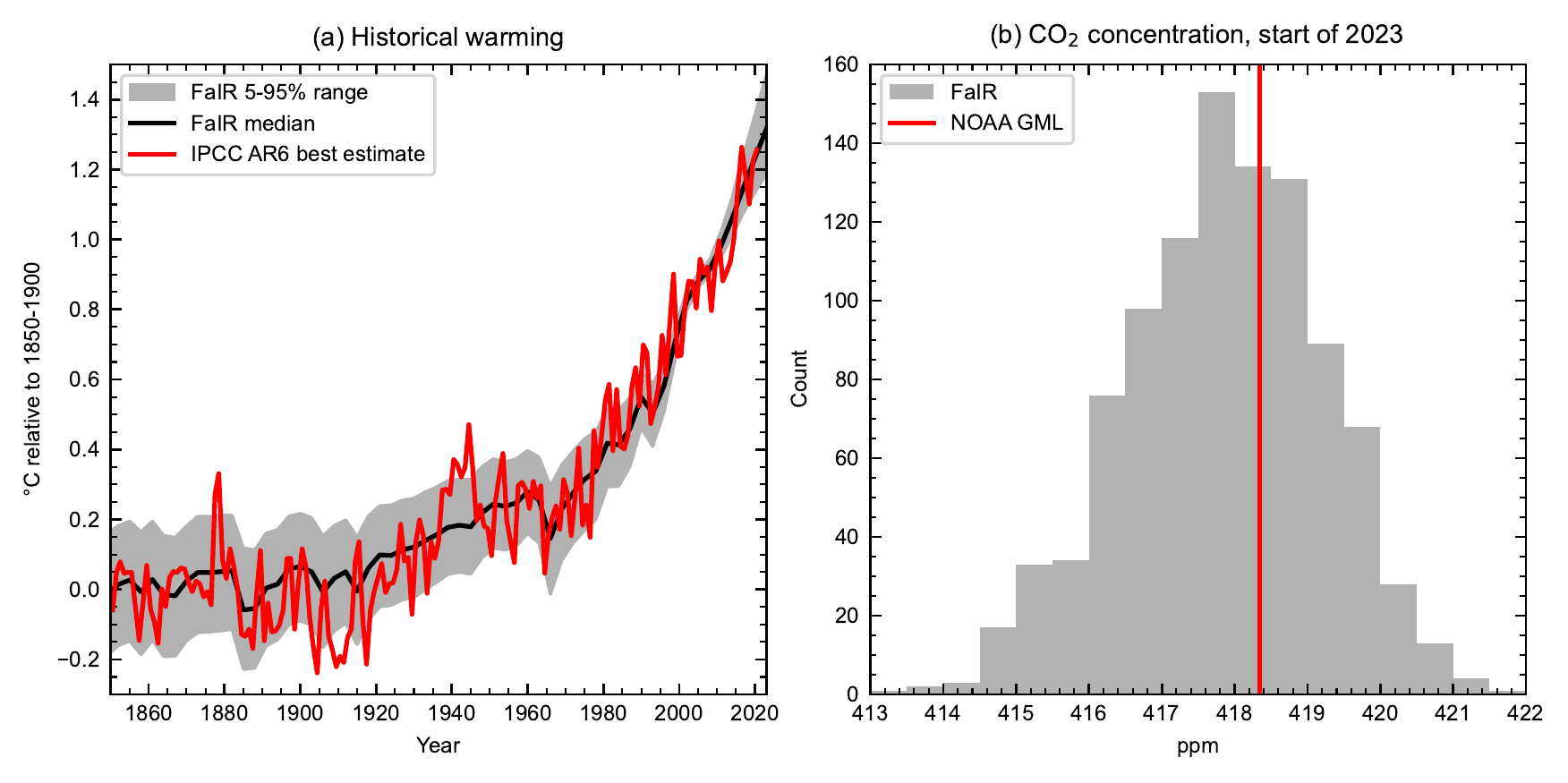}
\caption{(a) Historical global mean surface temperature in FaIR (5--95\% range in grey shading, median in black) compared to the IPCC's best estimate time series (red) from \citeauthor{Gulev2021} \cite{Gulev2021}. Temperatures use a baseline of 0.85°C above pre-industrial for 1995--2014, following IPCC. Widening spread near the beginning of the time series relates to observational uncertainty in present-day warming included in the ensemble. (b) Distribution of atmospheric CO$_2$ concentration at the start of 2023 from FaIR initialised in 1750 (grey histogram) compared to NOAA's global mean surface dataset (red line). Start of year 2023 CO$_2$ concentrations were estimated from extrapolating the 12-month trend value from December 2022 forward for half a month. Data was obtained from \url{https://gml.noaa.gov/webdata/ccgg/trends/co2/co2_mm_gl.txt} (accessed 3 April 2023).}
\label{fig:historical}
\end{figure*}

We produce a 1001 member posterior sample of FaIR parameters from a 1.5 million member prior ensemble. 
The 1001 ensemble members simultaneously span IPCC assessed ranges of ECS (e.g. 90\% of the distribution lying within 2--5°C), transient climate response (TCR), ocean heat content change from 1971--2018, global mean surface temperature from 1995--2014 relative to 1850--1900, aerosol effective radiative forcing (ERF; 2005--2014 relative to 1750), CO$_2$ concentrations in 2014 and future warming projected under SSP2-4.5 in 2081--2100 (Supplementary Material sect. 3). We verify that FaIR reproduces historical observed warming including its uncertainty (\cref{fig:historical}a) and present-day CO$_2$ atmospheric concentrations (\cref{fig:historical}b) when run with historical emissions from 1750 at a 3-year timestep.

As DICE only models CO$_2$ emissions, non-CO$_2$ emissions are treated as an external forcing so that the total forcing $F = F_{\mathrm{CO}_2} + F_{\mathrm{ext}}$. To generate $F_{\mathrm{ext}}$ for our scenarios we run FaIR offline using the 1001-member posterior ensemble under SSP2-4.5, SSP1-2.6 and SSP1-1.9 emissions. These three SSP scenarios are used in our study to model Nordhaus' ``welfare optimal'', 2°C and 1.5°C overshoot scenarios respectively for 1750--2500 \cite{Smith2023cal} (\cref{sec:scenarios}) which have similar warming levels to each SSPs. For 2023 onwards we export $F_{\mathrm{ext}}$ from each ensemble member and each scenario and use this as an exogenous input to the DICE runs. This captures uncertainty in the strength of non-CO$_2$ forcing, including aerosols, but not uncertainties in future emissions. We use GCP CO$_2$ emissions from 1750--2022 and non-CO$_2$ emissions from the RCMIP dataset \cite{Nicholls2020,Nicholls2021,Nicholls2021rcmip}. For CO$_2$, we harmonize \cite{Gidden2018} the CO$_2$ emissions to ensure a smooth transition between the GCP historical and the SSP future for CO$_2$.


FaIR v2.1.0 uses the \citeauthor{Meinshausen2020} \cite{Meinshausen2020} relationship of ERF from concentrations of CO$_2$, CH$_4$ and N$_2$O which includes radiative band overlaps between gases. As DICE only models CO$_2$ concentrations explicitly we revert to the logarithmic formula for CO$_2$ forcing \cite{Myhre1998}

\begin{equation}
F_{\mathrm{CO}_2} = F_{2\times\mathrm{CO}_2} \frac{\log(C_{\mathrm{CO}_2}/C_{\mathrm{CO}_2,\mathrm{ref}})}{\log 2}
\label{eq:forcing_co2}
\end{equation}
where $C_{\mathrm{CO}_2}$ is the CO$_2$ concentration in parts per million volume (ppm) and $C_{\mathrm{CO}_2,\mathrm{ref}}$ is the pre-industrial concentration. $F_{2\times\mathrm{CO}_2}$ is the ERF from a doubling of CO$_2$ above pre-industrial concentrations. To transition from the Meinshausen formula to the logarithmic formula we calculate an effective $F_{2\times\mathrm{CO}_2}$ from each historical ensemble member to use in the corresponding DICE simulation by rearranging \eqref{eq:forcing_co2} and using 2023 values of $F_{\mathrm{CO}_2}$ and $C_{\mathrm{CO}_2}$. 

For computing the temperature response to ERF, FaIR uses an impulse-response formulation of the well-known $n$-layer energy balance model \cite{Geoffroy2013}. We use $n=3$, expected to be sufficient to capture short- and long-term climate responses to forcing \cite{Leach2021,Cummins2020}. 
Results from the offline historical FaIR runs are saved out for 2023 and used as initial conditions for DICE. The temperatures of the three ocean layers in 2023 are re-baselined such that the uppermost layer (a proxy for global mean near-surface air temperature) is defined to be 0.85°C above pre-industrial over the 1995-2014 mean, this being the best estimate assessed warming in the IPCC AR6 WG1 \cite{Gulev2021} and following the treatment of scenario assessment in IPCC AR6 WG3 \cite{Byers2022,Riahi2022,Kikstra2022}. The other two ocean layers are adjusted by the same amount that was required to fix the uppermost layer at 0.85°C, maintaining relative differences.

FaIR uses four atmospheric boxes to model CO$_2$ concentrations (Supplementary Material sect. 2.1). The carbon mass in each box is also saved out of the historical run and used for initialising DICE in 2023. The sum of the atmospheric boxes (a mass anomaly above pre-industrial) and the pre-industrial mass (a probabilistic parameter sampled in ref. \cite{Smith2023cal}) gives the initial atmospheric CO$_2$ concentration at the start of 2023 (\cref{fig:historical}b). 

\subsection{Scenario construction}
\label{sec:scenarios}
The three scenarios (Nordhaus' ``optimal'', well-below 2°C and 1.5°C overshoot) are differentiated solely by their discount parameters and the SSP scenario chosen to represent their non-CO$_2$ forcing.

DICE uses Ramsey-style discounting \cite{Ramsey1928} to express future values in today's equivalents. The social discount rate $r$ is
\begin{equation}
    r = \rho + \eta g
\end{equation}
where $\rho$ is the pure rate of time preference, $\eta$ is the elasticity of marginal utility of consumption and $g$ is per-capita growth in consumption in percent. In Nordhaus' ``optimal'' scenario we use the default DICE-2016R parameters of $\rho = 1.5\%$ and $\eta = 1.45$ resulting in social discount rates around 3.1\%.  The 2°C scenario uses $\rho = 0.35\%, \eta = 0.35$ and the 1.5°C scenario uses $\rho = 0.12\%, \eta = 0.12$, resulting in very low social discount rates centred around 1.4\% and 0.6\% respectively. These parameters have been selected solely to achieve the goals of constructing scenarios that meet the Paris Agreement targets and are not necessarily constructed to be economically meaningful.

\section{Results}

\subsection{CO$_2$ emissions pathways}

\begin{figure*}[tbhp]
\centering
\includegraphics[width=17.8cm,height=17.8cm]{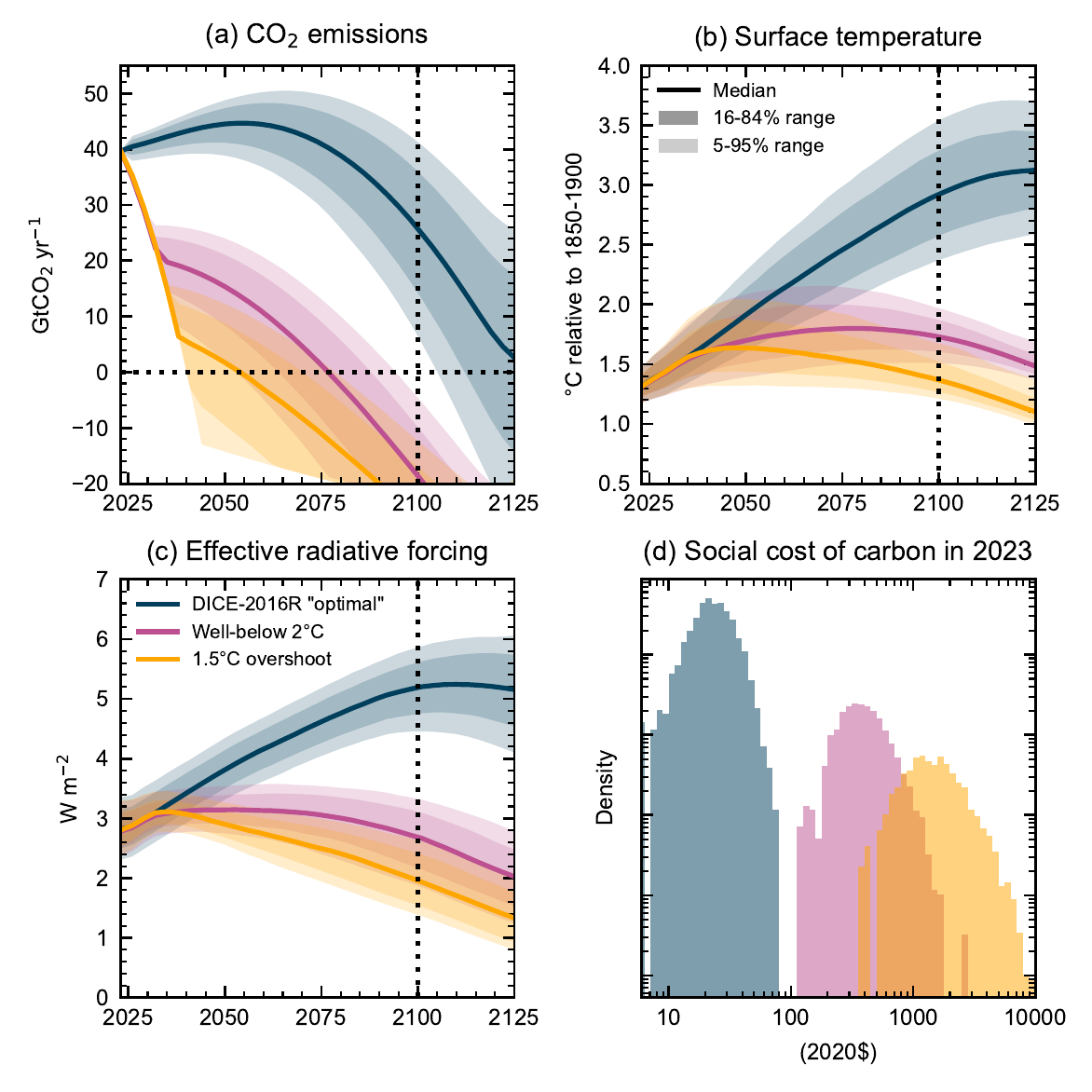}
\caption{Emissions, climate and economic projections for three scenarios using FaIR-DICE. (a) CO$_2$ emissions from energy and industrial processes for  the ``optimal'' (blue), 2°C (pink) and 1.5°C (yellow) scenarios. (b) Temperature projections. (c) Total effective radiative forcing projections. (d) Histogram of year-2023 SCC (in 2020\$) on a log-log scale. In (a-c), light shading shows 5-95\% range, darker shading shows 16-84\% range and solid lines show ensemble medians.}\label{fig:climate_projections}
\end{figure*}

\Cref{fig:climate_projections} shows the headline projections for the three scenarios, which are summarized in \cref{tab:results}. In each scenario, a wide range of allowable CO$_2$ emissions consistent with the ensemble warming classification are shown. The Nordhaus ``optimal'' pathway produces a level of total CO$_2$ emissions ranging from 5--41 Gt CO$_2$ yr$^{-1}$ in 2100 (5--95\% range), with a relatively smaller spread in 2050. In contrast, the 2°C and 1.5°C scenarios show larger spreads in their 2050 CO$_2$ emissions (2--24 and \textminus14 to +12 Gt CO$_2$ yr$^{-1}$ respectively). This suggests that climate uncertainty alone can either demand high levels of net negative emissions or permit substantial residual positive emissions in mid-century. By the end of the century, a majority of 1.5°C scenarios approach the maximum abatement level allowed in DICE (120\% of gross emissions), evidenced by the 5th and 50th percentile being at the same \textminus23 CO$_2$ yr$^{-1}$ level.

The observation that all 1.5°C and 2°C pathways follow the emissions abatement upper bound of $\mu(t) = 0.15t$ (emissions lower bound) for the first few periods (\cref{fig:climate_projections}a) demonstrates that decarbonizing as rapidly as possible in the near term is welfare-optimal under Paris Agreement long-term temperature constraints.

\begin{table*}
\centering
\caption{Key results from the three scenarios. All correlations are significant at the 1\% level.}
\label{tab:results}
\resizebox{\textwidth}{!}{
\begin{tabular}{lrrr}
Variable & Nordhaus ``optimal'' & Well below 2°C & 1.5°C low overshoot \\
\midrule
CO$_2$ emissions 2050 (Gt CO$_2$ yr$^{-1}$) & 45 (39--49) & 15 (2--24) & 2 (\textminus14 to +12) \\
CO$_2$ emissions 2100 (Gt CO$_2$ yr$^{-1}$) & 25 (5--41) & \textminus19 (\textminus23 to \textminus5)   & \textminus23 (\textminus23 to \textminus13) \\
Net zero CO$_2$ year         & 2129 (2105--2152) & 2077 (2053--2094) & 2054 (2040--2079) \\
Social cost of carbon 2023 (2020\$ (t CO$_2$)$^{-1}$) & 26 (15--44) & 439 (237--934) & 1759 (821--4434)\\
Peak warming (°C relative to 1850--1900)         & 3.1 (2.7--3.7) & 1.8 (1.5--2.2) & 1.6 (1.3--2.1) \\
Warming 2100 (°C relative to 1850--1900)          & 2.9 (2.4--3.6) & 1.7 (1.5--2.0) & 1.4 (1.2--1.7) \\
Effective radiative forcing 2100 (W m$^{-2}$)             & 5.2 (4.4--5.9) & 2.7 (1.9--3.3) & 1.9 (1.4--2.6) \\
ECS/SCC correlation coefficient & .51 & .74 & .74 \\
ECS/2050 CO$_2$ emissions correlation coefficient & \textminus.48 & \textminus.72 & \textminus.76 \\
2014 aerosol forcing/SCC correlation coefficient & \textminus.64 & \textminus.60 & \textminus.59 \\
2014 aerosol forcing/2050 CO$_2$ emissions correlation coefficient & .61 & .59 & .56 \\
Near-term discount rate (\%) & 3.1 (3.1--3.2) & 1.4 (1.2--1.6) & 0.6 (0.2--0.8) \\
\bottomrule
\end{tabular}}
\end{table*}

\subsection{Timing of net zero CO$_2$}
The 1.5°C scenario reaches net zero CO$_2$ emissions with an ensemble median year of 2054, which is consistent with the C1 scenario category of IPCC AR6 WG3. The well-below 2°C ensemble has a median net zero CO$_2$ emissions year of 2077, which is a little later than the IPCC's C3 scenario category. 
The ``optimal'' ensemble does not reach net zero CO$_2$ emissions this century, but does reach net zero with a median year of 2129. This demonstrates the utility of extending scenarios beyond 2100 to consider longer-term impacts.

\subsection{Global mean surface temperature}
Global mean surface temperature reaches 2.9°C above pre-industrial in the ``optimal'' pathway, peaking at 3.1°C in the 22nd century (\cref{fig:climate_projections}b). The 2°C and 1.5°C scenarios exhibit peak warming this century, consistent with net-zero CO$_2$ dates well before 2100. The 1.5°C overshoot ensemble has a peak warming of 1.6°C. As more than 33\% of the ensemble members have a peak warming above 1.5°C, this ensemble does not meet the IPCC definition of ``low overshoot'' (category C1 in \cite{Riahi2022}) and would fall into the C2 (1.5°C high overshoot) category. Indeed, it is difficult to avoid overshooting 1.5°C at all from today's starting level of warming, even under very rapid emissions phase-out scenarios \cite{Dvorak2022}. The 2°C scenario is within in the definition of C3 from the IPCC (67\% of the ensemble remaining below 2°C).

\subsection{Effective radiative forcing} The total median ERF (\cref{fig:climate_projections}c) in 2100 is 5.2 W m$^{-2}$ in the ``optimal'' scenario, 2.7  W m$^{-2}$ in the 2°C scenario and 1.9  W m$^{-2}$ in the 1.5°C scenario. Non-CO$_2$ forcing pathways were provided from SSP2-4.5, SSP1-2.6 and SSP1-1.9 respectively, though the total ERF is dominated by the CO$_2$ component. In the 2°C and 1.5°C scenarios, the median ERF in 2100 is very similar to the non-CO$_2$ scenario nameplate forcing in 2100. SSP1-2.6 and SSP1-1.9 were designed to be `well-below 2°C'' and 1.5°C-consistent scenarios respectively and our ERF results are therefore consistent with the SSP scenario framework  \cite{ONeill2016}.

\subsection{Social cost of carbon}
The SCC shows a wide uncertainty range for each scenario, with the spread increasing for stronger mitigation (lower discount rates) (\cref{fig:climate_projections}d). The 5--95\% uncertainty range is approximately a factor of three (15--44 \$ (t CO$_2$)$^{-1}$), four (237--934 \$ (t CO$_2$)$^{-1}$) and five (821--4434 \$ (t CO$_2$)$^{-1}$) for the ``optimal'', 2°C and 1.5°C cases respectively (values are reported in 2020 US dollars). Our findings that lower discount rates show more spread in the relative range of SCC when climate uncertainty is taken into account is consistent with \cite{Folini2022,Rennert2022}. Our hypothesis for this is that for higher discount rates, some of the long-term costs of climate damages are discounted away leading to a greater spread in long-term warming (\cref{fig:climate_projections}b), a lesser spread in near-term mitigation effort (\cref{fig:climate_projections}a), and hence a smaller spread in SCC.

\subsection{Relationships between climate sensitivity, aerosol radiative forcing and social cost of carbon}
\begin{figure}[tbhp]
\centering
\includegraphics[width=17.8cm,height=17.8cm]{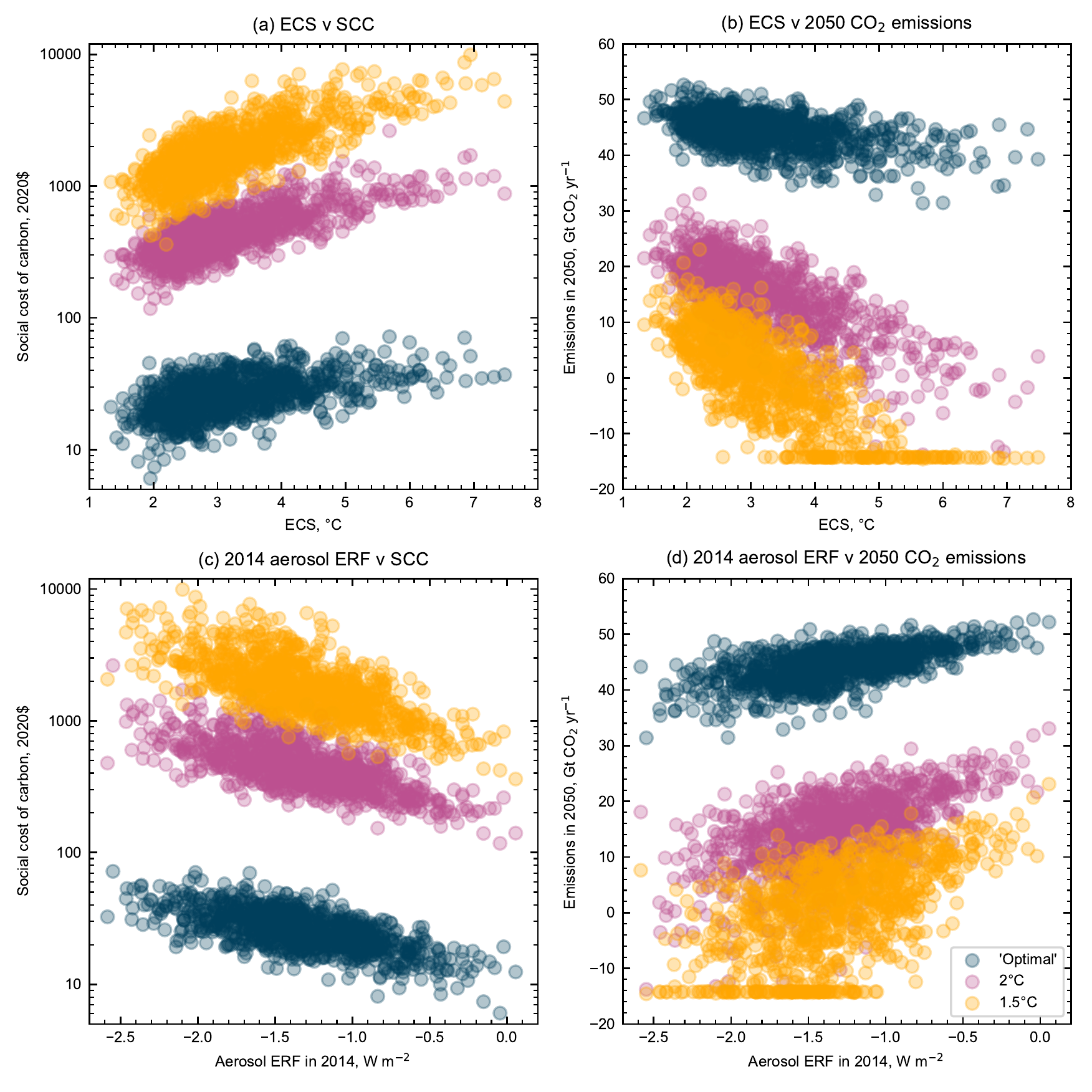}
\caption{Relationship between parameters. (a) ECS versus social cost of carbon; (b) ECS versus CO$_2$ emissions in 2050; (c) 2014 aerosol ERF versus social cost of carbon; (d) 2014 aerosol ERF versus CO$_2$ emissions in 2050.}\label{fig:ecs_emissions}
\end{figure}

There is a strong positive correlation between SCC and ECS \cite{Anderson2014}, particularly in 1.5°C and 2°C mitigation scenarios (\cref{fig:ecs_emissions}a). This follows from the fact that if climate sensitivity is high, emissions need to be abated more aggressively to maintain a similar warming level (and similar level of associated climate damages) compared to a case where climate sensitivity is low. Stronger abatement necessitates a higher social cost of carbon. This also confirms that reducing climate sensitivity uncertainty can lead to better informed estimates of the social cost of carbon and net present benefits \cite{Hope2015}.

The negative correlation between ECS and net CO$_2$ emissions in 2050 is shown in \cref{fig:ecs_emissions}b, showing that stronger emissions abatement is required if climate sensitivity is high as a corollary of the discussion above. In 2050, the maximal level of mitigation (net emissions of \textminus14 GtCO$_2$ yr$^{-1}$) is reached in several of the 1.5°C ensemble members. These tend to be clustered towards higher values of ECS, though moderate ECS between 3 and 4°C could still require very high levels of abatement. 

Alongside climate sensitivity, present-day aerosol ERF is a strong predictor of 21st century warming \cite{Smith2019,Watson-Parris2022}. In \cref{fig:ecs_emissions}c there is a negative correlation between aerosol ERF in 2014 and social cost of carbon, and in \cref{fig:ecs_emissions}d a positive correlation between aerosol ERF and 2050 CO$_2$ emissions. These are the opposite signs to the correlations related to ECS in \cref{fig:ecs_emissions}a-b, and is due to ECS and aerosol ERF being negatively correlated in observationally consistent climate simulations \cite{Smith2018}. A strong negative aerosol forcing is associated with a sensitive climate, as historical greenhouse gas warming has been offset by cooling aerosols. Aerosol forcing may be easier to constrain than ECS, and this indicates there are also net present economic benefits to reducing uncertainty in aerosol forcing \cite{Watson-Parris2022}.

\section{Discussion and conclusions}
We show that the optimal CO$_2$ emissions pathways and social cost of carbon are sensitive to physical climate uncertainty, including ECS and present-day aerosol forcing. Due to climate uncertainty alone, a range of CO$_2$ emissions pathways could be consistent with a 1.5°C future, ranging from requisition of a high level of net negative emissions to allowance of a substantial level of residual positive emissions. However, there are few plausible climate states forgiving enough to allow achieving Paris-compliant climate goals (well-below 2°C or 1.5°C) without net negative emissions in the second half of the century, evidenced by the emissions in the 95th percentile of the 2°C scenario being below zero in 2100 (\cref{fig:climate_projections}a). Net negative emissions in 2100 are at \textminus23 Gt CO$_2$ yr$^{-1}$ in more than half of the 1.5°C ensemble, this being the maximum abatement of 120\% of gross emissions assumed in DICE. We note that this level of net negative emissions may not be achievable in reality due to feasibility constraints \cite{Fuss2018,IPCC2022}.

There is a strong positive correlation between SCC and ECS, and negative correlation between aerosol forcing and ECS, where high climate sensitivity or strong aerosol forcing leads to aggressive abatement being socially optimal, and hence leads to a higher SCC. Owing to this, there is a relationship between climate sensitivity (or aerosol forcing) and emissions which can be contextualised as a climate-abatement feedback. This feedback is straightforward to demonstrate in DICE but is missing from PB-IAMs, at least when being used to construct emissions scenarios for IPCC \cite{Riahi2022} and policymaking. 

In PB-IAMs, there exists the opportunity to consider the processes under which climate change causes economic losses (or benefits). Climate change may lead to impacts on energy generation \cite{Cronin2018}, heating and cooling demand \cite{vanRuijven2019}, labour productivity, agriculture, bioenergy, and sea-level rise \cite{Rennert2022}, in addition to remedial costs resulting from climate catastrophes that will likely increase in severity and frequency \cite{Seneviratne2021}. While in some cases difficult, incorporation of these effects into PB-IAMs will lead to more realistic emissions scenarios, particularly in high emissions pathways where high levels of warming increases climate damages, reduces GDP and consumption, and hence is a negative feedback onto emissions \cite{Woodard2019}. 

Our ``optimal'' scenario has a lower median SCC at \$26 than DICE-2016R which is \$31 in 2015\$ (\$34 in 2020\$). This is despite the lower effective discount rate in our study (3.1\% versus DICE-2016R's 4.25\%), driven by lower near-term per-capita consumption growth rates. An updating and recalibration of the economic assumptions used in DICE partly accounts for the differences, particularly our lower future population projections compared to DICE-2016R (\cref{sec:dice}). The social discount rates required to construct our scenarios are significantly lower than those used in the literature for mitigation scenarios. Our 2°C scenario uses the same discount rate, by coincidence, as Stern's assessment of the costs of climate change \cite{Stern2007}. As the social discount rate relies on the growth in consumption, and consumption is affected by both by investment diverted towards emissions abatement and climate damages, the near-term discount rate is affected by climate uncertainty in our scenarios and is not a single value across all ensemble members (table \ref{tab:results}). Our analysis shows that meeting 1.5°C with limited overshoot would require a very high carbon price, with a median estimate of \$1759 (t CO$_2)^{-1}$ and 95th percentile of \$4434 (t CO$_2)^{-1}$. 

The social discount rate is one of the most contested and controversial parameters in climate economics \cite{Haensel2020}. Nordhaus \cite{Nordhaus2017} suggests the discount rate should be a continuation of the real risk-free interest rate in the recent past, and opts for a discount rate in DICE-2016R of 4.25\%. Stern \cite{Stern2007} argues that the discount rate is a subjective valuation of the welfare of future generations compared to the present, and is a normative choice, putting forward an ethical basis for lower discount rates \cite{Dietz2015}. A recent evaluation of the SCC for recommendation to US policymakers uses a preferred discount rate of 2\% \cite{Rennert2022}. Our use of the discount rate as a control dial on the acceptable level of future warming puts us more in the ``normative choice'' camp of Stern. Regardless of viewpoint, the fact that three very different scenarios are achievable by modifying the discount rate confirms that discounting is one of the most influential parameters controlling emissions pathways and social cost of carbon \cite{Anderson2014,Yang2018,Miftakhova2021}.

In every ensemble member, a cost-benefit optimal emissions pathway is constructed, with the assumption that in each of these 1001 different ``worlds'' the social planner knows the state of the climate system in advance. It is likely that as climate change unfolds over the coming decades, uncertainty in emergent parameters in the climate system such as the ECS will reduce; we will simply have more observational evidence to draw upon \cite{Hope2015}. This reduction in uncertainty or updating of knowledge over time would be a useful future analysis. Another additional avenue of future study is the relative contributions of socioeconomic (e.g. growth in population, carbon intensity, total factor productivity, discount rate) and climate uncertainties on total variation in social cost of carbon and emissions pathways, including their time dependence. Although we include their forcing contributions and uncertainties and report on the dependency of SCC on aerosol ERF, non-CO$_2$ emissions are not calculated endogenously. Doing so from a process perspective would require modelling of cost-abatement curves in several sectors and substantially increase the complexity of the analysis, but relationships between key important non-CO$_2$ forcers and fossil CO$_2$ could be sought from a large database of PB-IAM scenarios \cite{Rogelj2014,Lamboll2020} at a relatively low computational cost, as we do for land-use CO$_2$. Notwithstanding its simplicity, this study highlights the importance of incorporating climate uncertainty into IAM-derived emissions scenarios.

\section*{Acknowledgments}
CJS was supported by a NERC-IIASA Collaborative Research Fellowship (NE/T009381/1) and Horizon Europe project WorldTrans (101081661). AAK was supported by the Engineering and Physical Sciences Research Council, United Kingdom, grant/award no. EP/P022820/1. PY was supported by the Climate Compatible Growth programme, which is funded by UK aid from the UK government. The views expressed herein do not necessarily reflect the UK government’s official policies. DF was supported by the Swiss National Science Foundation (SNF) under project ID `Can Economic Policy Mitigate Climate-Change?'. For the purpose of open access, the author has applied a Creative Commons Attribution (CC BY) license to any Author Accepted Manuscript version arising. The authors declare no conflicts of interest.

\section*{Data availability statement}
The data that support the findings of this study are openly available from \url{https://doi.org/10.5281/zenodo.8155285} \cite{Smith2023dice}.

\bibliography{main}

\end{document}


\maketitle

\section{Modifications to DICE-2016R}

Our starting point for analysis in this paper is the DICE-2016R integrated assessment model (IAM). DICE-2016R is a well-known cost-benefit IAM and is well-documented in the literature \cite{Nordhaus2013,Nordhaus2017}. Recently, a beta version of DICE-2023R has been submitted as an NBER report which incorporates FaIR v1.0 as its carbon cycle and climate module components \cite{Barrage2023}. We comment extensively on DICE-2023R in \cref{sec:20232016}, and compare both DICE-2016R and DICE-2023R to the version presented in this study which we name FaIR-DICE.

\subsection{Time step and time horizon}

We reduce the model time step in DICE-2016R from 5 years to 3 years. The benefits are that this allows for an integral number of periods from the nominal pre-industrial start year to the present-day (2023 minus 1750 is divisible by 3), while maintaining an integral number of periods to 2050 (2050 minus 1750 is also divisible by 3). It also allows model-derived mitigation trajectories to be more responsive. This is important given the proximity to 1.5°C in the present day and limited headroom available to stay under this limit.  

\subsection{Population}

The projections of world population from DICE-2016R are updated with the median projection of 10,000 scenarios from the Resources For the Future Socioeconomic Pathways (RFF-SPs) \cite{Rennert2022,Raftery2023}.

\begin{figure*}[tbhp]
\centering
\includegraphics[width=8.9cm,height=8.9cm]{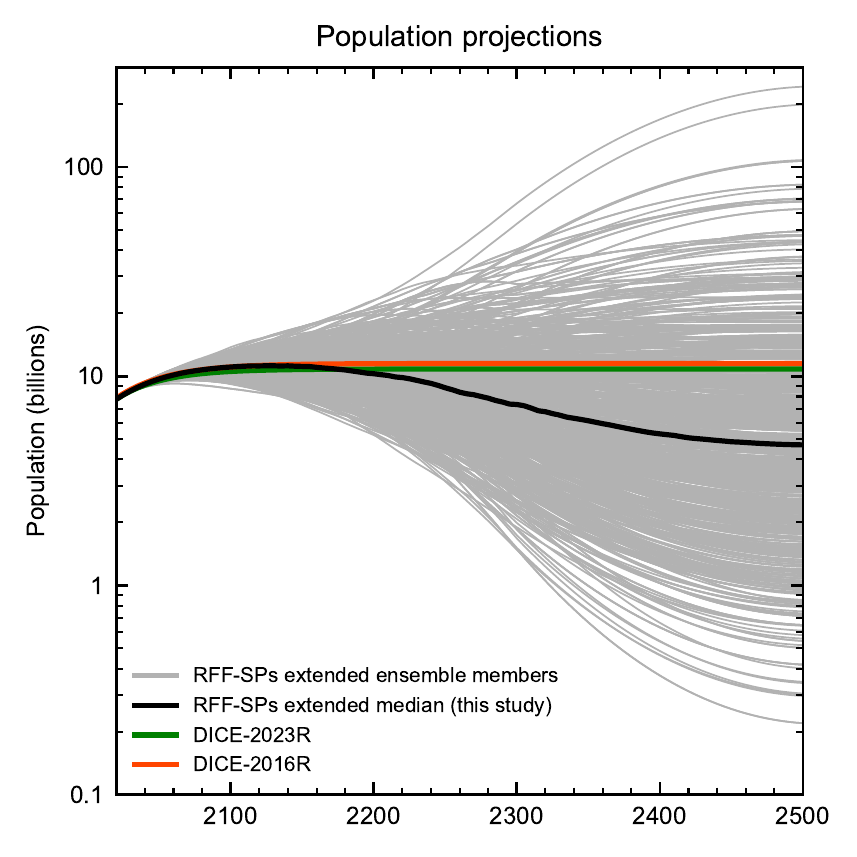}
\caption{Population projections from the 10,000 RFF-SPs from 2020 to 2300, extended to 2500 using the average growth rate of the last fifty years, declining to zero growth over 2300--2500 (coloured lines). The thick black line is the ensemble median, used for the population projection in this study.}
\label{fig:population}
\end{figure*}

The RFF-SPs run to 2300, which we extend to 2500 by taking the average growth rate over 2250--2300 in each projection and linearly declining this growth rate to zero by 2500. This produces a very wide range of potential population scenarios ranging from 200 million to 200 billion people in 2500 (\cref{fig:population} grey lines). The median pathway used for our ensemble is depicted in \cref{fig:population} by the black curve.

In the median of the RFF-SP projections, global population peaks in 2116 at 11.2bn, declining to 7.3bn in 2300 and 4.9bn in 2500. From 2200 onwards this population trajectory is substantially different to DICE-2016R (orange line) and DICE-2023R (green line), which assume asymptotic convergences to 11.5bn and 10.8bn respectively by 2500.

\subsection{Model for future land-use CO$_2$ emissions}
\label{sec:landuse}

The DICE-2016R and DICE-2023R models use exogenous time series of CO$_2$ emissions from agriculture, forestry and other land use (AFOLU). These emissions do not result from fossil fuel combustion and have differing economic and social drivers to fossil fuel emissions that are not modelled explicitly in DICE. Nevertheless, the AFOLU sector will play a vital role in climate change mitigation, and land-based carbon dioxide removal strategies (afforestation and reforestation) are featured heavily in process-based (PB-) IAM scenarios that limit global mean temperatures to 1.5°C or 2°C \cite{statecdr2023}. Therefore, it could be reasonably expected that a mitigation scenario that took strong action on fossil emissions would also address the AFOLU sector, even if the causal drivers are not strongly linked. 

In this regard, we seek a regression relationship predicting CO$_2$ AFOLU emissions $E_{\mathrm{AFOLU}}$ from FFI emissions $E_{\mathrm{FFI}}$ and period $t$, derived from 1202 PB-IAM scenarios from the IPCC WG3 database from 2030 to 2100 \cite{Byers2022}:

\begin{equation}
E_{\mathrm{AFOLU}} = \alpha + \beta_1 E_{\mathrm{FFI}} + \beta_2 t.
\label{eq:afolu}
\end{equation}

\begin{figure}
     \centering
     \begin{subfigure}[t]{8.9cm}
         \centering
         \includegraphics[width=\textwidth]{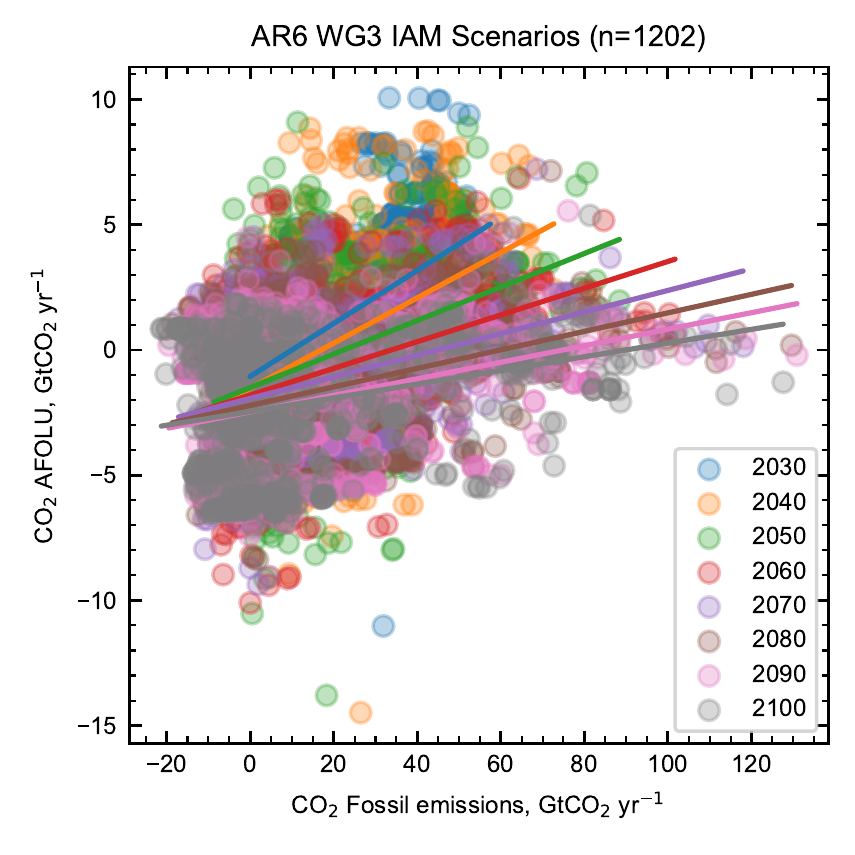}
         \caption{Comparison of $E_{\textrm{AFOLU}}$ with $E_{\textrm{FFI}}$ for 1202 PB-IAM scenarios submitted to the IPCC AR6 WG3 database. Regression lines drawn at 10 year intervals from 2030 to 2100.}
         \label{fig:afolu_reg}
     \end{subfigure}
     \hfill
     \begin{subfigure}[t]{8.9cm}
         \centering
         \includegraphics[width=\textwidth]{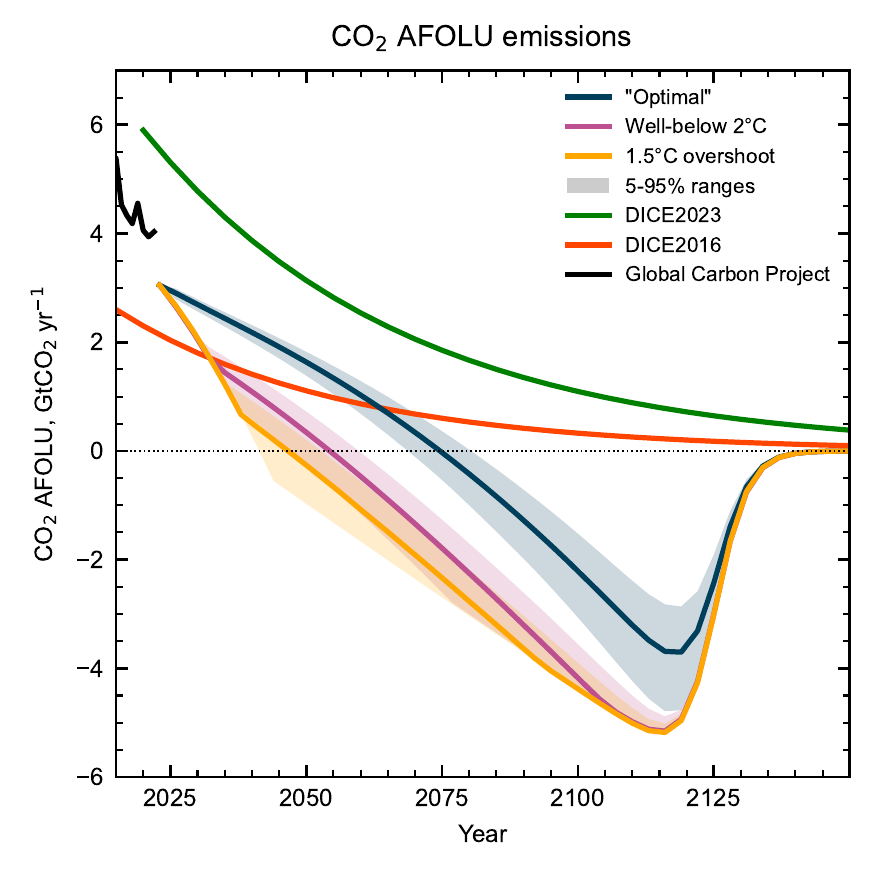}
         \caption{Comparison of $E_{\textrm{AFOLU}}$ from the three scenario ensembles in this study (coloured lines and uncertainty ranges) with DICE-2016 (dotted) and DICE-2023 (dashed) exogenously supplied time series. Historical best estimate emissions from Global Carbon Project are shown in black as a comparison.}
         \label{fig:afolu_path}
     \end{subfigure}
        \caption{AFOLU emissions derivation for the scenario pathways in this study.}
        \label{fig:afolu}
\end{figure}

\Cref{fig:afolu_reg} shows there is a positive, time-dependent relationship between fossil emissions and AFOLU emissions from PB-IAM scenarios justifying the form of \cref{eq:afolu} ($r^2=.47$). We derive $\alpha=1.54$ GtCO$_2$ yr$^{-1}$, $\beta_1 = 0.0464$, $\beta_2 = -0.189$ GtCO$_2$ yr$^{-1}$ period$^{-1}$. \Cref{eq:afolu} is only defined to 2100. To extend to 2500, we follow the assumption of a phase down to zero CO$_2$ AFOLU emissions by 2150 used for the SSP scenario extensions in \cite{Meinshausen2020}. This is achieved by multiplying \cref{eq:afolu} by a logistic curve $l(t)$ centred around $t=35$ (model year 2125):

\begin{equation}
l(t) = \left(1 - \frac{1}{1+e^{-(t-35)}}\right).
\label{eq:phaseout}
\end{equation}

The comparison of our ensembles to the exogenous CO$_2$ AFOLU emissions in DICE-2016R and DICE-2023R is shown in \cref{fig:afolu_path}. All of our scenarios have negative AFOLU emissions during the later 21st century, which is typical of PB-IAM scenarios (\cref{fig:afolu_reg}). This is in contrast to the DICE time series where AFOLU emissions converge to zero from above. The impact of the logistic phase out can be seen in the 2100--2150 period. Our year-2023 CO$_2$ AFOLU emissions value, derived solely from the regression relationship, is also slightly closer to the historical best estimate from the Global Carbon Project (GCP; black line near the left axis in \cref{fig:afolu_path}) than in DICE-2023, though it should be cautioned that the 2022 land use emissions from GCP is preliminary, and land use emissions have large uncertainty.

\section{Reduced version of FaIR v2.1.0 in DICE}

The Finite-amplitude Impulse Response (FaIR) model is a reduced-complexity climate model used to project global mean surface temperatures from emissions of greenhouse gases and short-lived climate forcers, and other anthropogenic and natural influences on the climate system \cite{Leach2021,Smith2018,Millar2017fair}. We implement a reduced version of FaIR in DICE for future projections (2023--2500) that is based on FaIR v2.1.0. 
A model description paper for FaIR v2.1.0 is not yet in the scientific literature but it is similar to v2.0.0 \cite{Leach2021}. The specific implementation for DICE is described below.

DICE calculates CO$_2$ emissions from fossil fuel and industry $E_{\mathrm{FFI}}$ from economic activity, abatement level and energy intensity (eq. 2, main manuscript). Our treatment of CO$_2$ emissions from agriculture, forestry and other land use (AFOLU) are described in \cref{sec:landuse}. FaIR-DICE, based on DICE-2016R, does not model any other anthropogenic greenhouse gas or short-lived climate forcer explicitly. Therefore our reduced version of FaIR is limited to the carbon cycle module and the temperature response module (both described below) with non-CO$_2$ forcing introduced exogenously (described in the next section).

\subsection{Carbon cycle model}

We retain the four time-constant model of atmospheric CO$_2$ decay to a pulse emission from FaIR v2.0.0. Written in forward difference mode, this is

\begin{equation}
    R_{i}(t+1) = a_i E_{\textrm{FFI}}(t) \alpha(t) \tau_i \left(1 - \exp\left(-\frac{\Delta t}{\alpha \tau_i}\right)\right) + R_i(t) \exp\left(-\frac{\Delta t}{\alpha \tau_i}\right)
\end{equation}
where $R_i$ is the carbon content in pool $i$, $\Delta t$ is the model timestep (3 yr), $a_i$ is the amount of fresh emissions allocated to each pool, $\tau_i$ is the e-folding lifetime of each pool, and $\alpha$ is the lifetime scaling factor that models the effect of carbon cycle feedbacks. The values of $a_i$ and $\tau_i$ are derived from Earth System and intermediate complexity models in \cite{Joos2013} (\cref{tab:boxes}). In FaIR v2.1.0 a user-definable $\Delta t$ is incorporated, partly for applications including IAM coupling. Previously, the model timestep was hard-coded to be one year.

\begin{table*}
\centering
\caption{Partition fractions and lifetimes of the four carbon box impulse-response model in FaIR}
\label{tab:boxes}
\begin{tabular}{lrr}
Pool $i$ & Partition fraction $a_i$ & Lifetime $\tau_i$ (yr) \\
\midrule
1 & 0.2173 & $10^9$ (essentially $\infty$) \\
2 & 0.2240 & 394.4 \\
3 & 0.2824 & 36.54 \\
4 & 0.2763 & 4.304 \\
\bottomrule
\end{tabular}
\end{table*}

Carbon uptake efficiency reduces with increasing warming and accumulated CO$_2$ in land and ocean sinks \cite{Revelle1957,Friedlingstein2006}. In FaIR, the increase in atmospheric lifetime of CO$_2$ is parameterised by $\alpha$. $\alpha$ is determined from the 100-year time-integrated airborne fraction from a pulse of CO$_2$ emissions, $I_{100}$:
\begin{equation}
\label{eq:cc_feedback}
    \alpha(t) = g_0 \exp\left(\frac{I_{100}(t)}{g_1}\right)
\end{equation}
where the airborne fraction is the proportion of CO$_2$ remaining in the atmosphere following a pulse emission, and follows the empirically derived equation \cite{Millar2017fair}
\begin{equation}
\label{eq:iirf}
    I_{100}(t) = r_0 + r_u (1-f_a(t)) G(t) + r_t T_1(t) + r_a f_a(t) G(t).
\end{equation}

In \cref{eq:iirf}, $r_0$ is the pre-industrial value of $I_{100}$, $r_u$ is the sensitivity of $I_{100}$ to accumulated carbon in land and ocean sinks, $r_t$ is sensitivity of $I_{100}$ to global mean surface temperature anomaly $T_1$, and $r_a$ is the sensitivity of $I_{100}$ to airborne CO$_2$. $G$ represents cumulative CO$_2$ emissions since pre-industrial and $f_a$ is the airborne fraction of CO$_2$ since pre-industrial (total increase of CO$_2$ in the atmosphere divided by $G$).  $I_{100}$ has units of yr.

In \cref{eq:cc_feedback}, $g_0$ and $g_1$ are normalisation constants defined as

\begin{equation}
    g_1 = \sum_i a_i \tau_i \left( 1 - \left( 1 + \frac{H}{\tau_i} \right) \exp\left(-\frac{H}{\tau_i} \right) \right)
\end{equation}

\begin{equation}
    g_0 = \exp \left(-\frac{\sum_i a_i \tau_i \left( 1 - \exp \left(-\frac{H}{\tau_i} \right) \right)}{g_1} \right)
\end{equation}
with $H = 100$ yr. The atmospheric CO$_2$ concentration $C_{\mathrm{CO}_2}$ is the sum of the four boxes added to the pre-industrial CO$_2$ atmospheric concentration ($C_{\mathrm{CO}_2,\mathrm{ref}}$), i.e.
\begin{equation}
    C_{\mathrm{CO}_2}(t) = C_{\mathrm{CO}_2,\mathrm{ref}} + \sum_i R_i(t).
\end{equation}

\subsection{Effective radiative forcing}

The total effective radiative forcing (ERF) $F$ is given by
\begin{equation}
F(t) = F_{\mathrm{CO}_2}(t) + F_{\mathrm{ext}}(t)
\end{equation}
where $F_{\mathrm{CO}_2}$ is the CO$_2$ component and $F_{\mathrm{ext}}$  is the and the exogenous non-CO$_2$ component. In this reduced version, the CO$_2$ forcing is computed from concentrations as \cite{Myhre1998}
\begin{equation}
\label{eq:forcing_co2}
F_{\mathrm{CO}_2}(t) = F_{2\times\mathrm{CO}_2} \frac{\log(C_{\mathrm{CO}_2}(t)/C_{\mathrm{CO}_2,\mathrm{ref}})}{\log 2}.
\end{equation}

In \cref{eq:forcing_co2}, $F_{2\times\mathrm{CO}_2}$ is the effective radiative forcing from a doubling of CO$_2$ above pre-industrial concentrations with a best estimate value of 3.93 W m$^{-2}$ \cite{Forster2021}. More complex relationships exist for CO$_2$ forcing that incorporate band overlaps between different greenhouse gases \cite{Etminan2016,Meinshausen2020}, however, as concentrations of other gases are not book-kept in DICE, the simpler formula in \cref{eq:forcing_co2} is used in FaIR-DICE.

\subsection{Temperature response module}

For computing the temperature response to effective radiative forcing, FaIR uses an impulse-response formulation of the well-known $n$-layer energy balance model \cite{Geoffroy2013}. The implementation here very closely follows the model of \citeauthor{Cummins2020} \cite{Cummins2020}. FaIR v2.1.0 allows for autocorrelated stochastic internal variability in temperature and forcing following \cite{Cummins2020}, but we do not use this in the present study. There is evidence that inclusion of internal variability increases the social cost of carbon (SCC) \cite{Calel2020} and this would be a valuable additional study in this setup.

We use $n=3$ layers, expected to be sufficient to capture short- and long-term climate responses to forcing \cite{Leach2021,Cummins2020}. The impulse-response model in forward timestepping mode can be written as
\begin{equation}
\mathbf{T}(t+1) = \mathbf{A_dT}(t) + \mathbf{b_d}F(t)
\end{equation}
where $\mathbf{T} = [T_1(t), T_2(t), T_3(t)]$ is a column vector of temperature anomalies representing the near-surface ocean layer and atmosphere, middle ocean, and deep ocean respectively. $T_1$ is used as global mean surface temperature anomaly throughout this study.

$\mathbf{A_d}$ and $\mathbf{b_d}$ are computed using the discretisation method of \citeauthor{Cummins2020} \cite{Cummins2020} with one modification in that we discard the first column and top row of $\mathbf{A_d}$ and first entry of $\mathbf{b_d}$ as calculated in \cite{Cummins2020}, as we do not use stochastic variability in forcing or temperature. The reader is referred to sections 4a and 4b of \citeauthor{Cummins2020} \cite{Cummins2020} for the calculation. The \citeauthor{Cummins2020} method takes a matrix of energy balance model parameters $\mathbf{A}$ where

\begin{equation}\mathbf{A} = \frac{1}{\Delta t}
\begin{bmatrix}
-\gamma \Delta t & 0 & 0 & 0 \\
1/C_1 & -(\kappa_1+\kappa_2)/C_1 & \kappa_2/C_1 & 0\\
0 & \kappa_2/C_2 & -(\kappa_2 + \epsilon \kappa_3)/C_2 & \epsilon \kappa_3/C_2 \\
0 & 0 & \kappa_3/C_3 & -\kappa_3/C_3
\end{bmatrix}
\label{eq:eb_matrix}
\end{equation}
and a column vector $\mathbf{b} = [\gamma, 0, 0, 0]$ as inputs. Here, $\kappa_j$ [W m$^{-2}$ K$^{-1}$] are heat transfer coefficients between adjacent ocean layers (with $-\kappa_1$ equivalent to the climate feedback parameter more commonly denoted as $\lambda$ in climate science literature), $C_j$ are heat capacities of each ocean layer [W m$^{-2}$ yr K$^{-1}$], and $\epsilon$ is the deep ocean efficacy parameter incorporated to simulate a forced pattern effect present in many Earth system models \cite{Held2010,Winton2010}. $\gamma$ is a parameter describing autocorrelation in the radiative forcing time series but is redundant if internal variability is not used. These parameters are easily calibrated to Earth System models (ESMs) as demonstrated in \cite{Geoffroy2013,Cummins2020}. 

\section{Constrained probabilistic ensemble of FaIR v2.1.0}

A semi-automated software package \emph{fair-calibrate} exists for producing calibrated, constrained probabilistic ensembles from the FaIR model (starting from FaIR v2.1.0). A separate manuscript is in preparation describing this process in detail. We provide an overview relevant to results in this paper here. We use the procedure and parameter values from v1.0.2 of \emph{fair-calibrate} which is available from \cite{Smith2023cal}. This reference also contains the calibration code which is fully reproducible. The calibration procedure produced uses the full standalone (rather than reduced, DICE-coupled) FaIR v2.1.0 model, is run at a one-year rather than three-year time step, and has internal variability switched on.  

\subsection{Calibration}

\subsubsection{Climate response}

The parameters of the three-layer energy balance model ($\kappa_j$, $C_j$, $\epsilon$ and $\gamma$, along with the ERF from a quadrupling of CO$_2$, $F_{4\times\mathrm{CO}_2}$) are calibrated to the response of 49 \emph{abrupt-4xCO2} ESMs participating in the Coupled Model Intercomparison Project Phase 6 (CMIP6) using the maximum likelihood method of \citet{Cummins2020}. We use the first 150 years from the quadrupled CO$_2$ run for consistency, even for models where more years are available. This provides us with a set of nine climate response parameters from 49 ESMs. 

\subsubsection{Carbon cycle feedbacks}

The parameters dictating the carbon cycle feedbacks are $r_0$, $r_u$, $r_t$ and $r_a$ defined in \cref{eq:iirf}. The parameters are calibrated from 11 carbon-cycle enabled Earth System models from CMIP6 using \emph{1pctCO2} and \emph{1pctCO2-rad} experiments as described in \cite{Leach2021}.

\subsubsection{Non-CO$_2$ forcing}

The full FaIR model that is used for the historical spin up also uses CMIP6 model results to calibrate the effective radiative forcing from aerosol-cloud interactions (ERFaci) to emissions of precursor species in 11 CMIP6 models (a similar method to \cite{Smith2021JGRA}), the sensitivities of chemically active species to ozone formation and destruction in six CMIP6 models (from \cite{Skeie2020}), and the sensitivities of chemically active species to methane lifetime in four CMIP6 models (from \cite{Thornhill2021erf,Thornhill2021fb}). There are many more anthropogenic and natural components that influence non-CO$_2$ forcings, with these sensitivities based on assessments from the IPCC Sixth Assessment Report (AR6) rather than CMIP6 models \cite{Forster2021,Smith2021AR6}.

\subsection{Sampling}
\label{sec:sampling}

We use a prior ensemble size of 1,500,000 parameter sets based on CMIP6 model calibrations or AR6 distributions, and use these parameter sets to run the full version of FaIR offline (i.e. uncoupled to DICE) from 1750--2100 under the historical + SSP2-4.5 scenario. Emissions of 53 greenhouse gas and short-lived climate forcers under historical + SSP2-4.5 are provided by the Reduced Complexity Model Intercomparison Project (RCMIP), with data available from \cite{Nicholls2021rcmip}. For short-lived climate forcers (SO$_2$, BC, OC, NH$_3$, CO, NMVOC, NOx), RCMIP uses the same input emissions sources prepared for running CMIP6 models, whereas greenhouse gases emissions are based on respected emissions inventories for CH$_4$ and N$_2$O and estimated from inversions of concentrations for halogenated greenhouse gases. We do not use the CO$_2$ FFI and CO$_2$ AFOLU emissions from RCMIP for the historical period up to 2021, instead using data from Global Carbon Project \cite{Friedlingstein2022} which is more up-to-date and ensures the beginning of the DICE projection period uses a more recent estimate of CO$_2$ emissions. To transition to the SSP2-4.5 scenario, the future emissions are harmonized \cite{Gidden2018} to the updated GCP historical  to ensure a smooth transition from the updated baseline to the future scenario.

\subsubsection{Climate response}

The nine energy balance model parameters are sampled from a multi-dimensional kernel density estimate function that is generated from the 49 CMIP6 models used in the calibration.

\subsubsection{Aerosol-cloud interactions}
\label{sec:aci}

There are four parameters that are used to model the ERFaci in FaIR, which are the sensitivity to SO$_2$, BC, and OC emissions, and a scale factor. The shape parameters are sampled from correlated kernel density estimates that are constructed by the fits to 11 CMIP6 models. The scale parameter is selected such that the resulting ERFaci in each ensemble member for the 2005--2014 period relative to 1750 is uniform in the range $-2$ to 0 W m$^{-2}$. 

\subsubsection{Aerosol-radiation interactions}
\label{sec:ari}

A number of species are involved in contributing to the ERF from aerosol-radiation interactions (ERFari) \cite{Szopa2021}. Aerosol formation is dominated by sulfate (primary source SO$_2$ emissions), BC, OC and nitrate (dominant source NH$_3$ emissions), with minor contributions from NOx, NMVOC, CH$_4$, N$_2$O, and halogenated greenhouse gases. 

A forcing efficiency for each species is calculated by dividing its contribution to ERFari as assessed by IPCC AR6 \cite{Szopa2021} by its change in emissions (or concentrations for greenhouse gases) between 1750 and 2019. The headline assessment of ERFari is $-0.3$ W m$^{-2}$ from 1750 to 2005--2014, which is reproduced using these forcing efficiencies (the assessed ERFari in 2019 was $-0.22$ W m$^{-2}$ owing to reductions in emissions). To sample uncertainties, we use a uniform distribution for the forcing efficiency of each species that ranges from zero to a factor of two times the best estimate. The addition of nine uniform random variables leads to an overall prior distribution of ERFari that is approximately trapezoidal.

\subsubsection{Carbon cycle feedbacks}

The carbon cycle feedback strengths are dictated by the $r_0$, $r_u$, $r_t$ and $r_a$ parameters in \cref{eq:iirf}. A four-dimensional kernel density estimate using the 11 model CMIP6 model calibrations in \cite{Leach2021} (their table 3) were used to sample from. 

\subsubsection{Ozone forcing}
\label{sec:ozone}

The starting point for sampling ozone forcing is the IPCC AR6 assessment of total (tropospheric plus stratospheric) ozone forcing from 1750 to 2019 of +0.47 W m$^{-2}$. This is based on the six-model mean historical time series from CMIP6 models in \cite{Skeie2020}. A correction for historical warming is applied to this time series using CMIP6 data of $-0.037$ W m$^{-2}$ K$^{-1}$ \cite{Thornhill2021fb} which is backed out of the forcing time series; this is because warming produces more water vapour, which produces more OH radicals, which are a chemical sink for ozone. Ozone is produced from chemical reactions that are facilitated from CH$_4$, NOx, N$_2$O, NMVOC and CO, and destroyed by halogenated greenhouse gases. The contribution of each factor to total ozone (in W m$^{-2}$) is derived from six CMIP6 models in \cite{Thornhill2021erf} for the 1850--2014 period which is then scaled up for each species to match the overall 1750--2019 assessment from AR6. The uncertainty in the ozone forcing contribution from each specie is also taken from \cite{Thornhill2021erf} and scaled. These values are converted to a forcing efficiency for each specie, as is done for ERFari. A Gaussian distribution is assumed for the forcing efficiency of each parameter.

\subsubsection{Effective radiative forcing uncertainty}

IPCC AR6 considered the effective radiative forcing in a few aggregated categories, and provided uncertainty ranges for each \cite{Forster2021} (see \cref{tab:forcing_scaling}). For many categories we used the same uncertainty ranges as AR6, which are applied as a scaling factor to the historical and future forcing time series (\cref{tab:forcing_scaling}). For solar forcing, the amplitude of the solar variability around a mean value is scaled by a factor, and an underlying trend over 1750 to 2019 is added. For ozone and aerosols, we do not use the AR6 uncertainties as priors but the priors described in the various preceding sections, which are similar to or wider than the corresponding AR6 assessments.

\subsubsection{CO$_2$ initial concentrations}

The pre-industrial concentrations of many greenhouse gases are uncertain. As we use present-day CO$_2$ concentrations as a constraint, we consider the possibility that the atmospheric concentration of CO$_2$ in 1750 was slightly higher or lower than the IPCC's best estimate. Clearly, as we do not have in-situ observations from this time, the uncertainty in concentrations is greater than in the present day. We use the AR6 assessment, drawing samples from a Gaussian distribution with mean 278.3 ppm and 90\% range 2.9 ppm \cite{Gulev2021}. Variation in pre-industrial CO$_2$ values affects both the carbon cycle and the CO$_2$ forcing calculation (\cref{eq:forcing_co2}). 

\begin{table*}
\centering
\caption{Effective radiative forcing scaling factor distributions used in the prior parameter sampling for FaIR. Symmetric uncertainty ranges are Gaussian, asymmetric ranges are half-Gaussian where two separate distributions are used for values below and above the best estimate. The solar forcing trend is Gaussian around +0.01 W m$^{-2}$. Some forcers do not take their uncertainties from AR6 and use different prior distribution uncertainties given by the corresponding section number.}
\label{tab:forcing_scaling}
\begin{tabular}{ll}
Forcer & 90\% uncertainty around best estimate, or source \\
\midrule
CO$_2$ & $\pm 12\%$  \\
CH$_4$ & $\pm 20\%$  \\
N$_2$O & $\pm 14\%$  \\
Halogenated GHGs & $\pm 19\%$ \\
Stratospheric water vapour & $\pm 100\%$ \\
Ozone & \cref{sec:ozone} \\
Aerosol-radiation interactions & \cref{sec:ari} \\
Aerosol-cloud interactions & \cref{sec:aci} \\
Aviation contrails & $-66 \%$ to $+70 \%$ \\
Black carbon on snow & $-100 \%$ to $+125 \%$ \\
Land use and irrigation & $\pm 50\%$ \\
Solar forcing amplitude & $\pm 50\%$ \\
Solar forcing trend 1750--2019 & $-0.06$ to $+0.08$ W m$^{-2}$ (Gaussian) \\
Volcanic forcing & $\pm 25\%$ \\
\bottomrule
\end{tabular}
\end{table*}

\subsection{Constraining}

From the 1,500,000 member prior ensemble we seek to produce a smaller constrained ensemble of parameters that fit both observational constraints and IPCC assessments. The prior ensemble produces a very wide range of climate projections, most of which are implausible and not useful for using for future projections (\cref{fig:prior}).

\subsubsection{First step: root-mean-square comparison to historical warming}
\label{sec:rmse}
The first constraining step uses a root-mean-square error (RMSE) comparison between the time series of historical temperature observations from the IPCC assessment \cite{Gulev2021} and each FaIR ensemble member. Ensemble members that differ from the historical time series with a RMSE of more than 0.16 K are rejected. This rules out 92\% of the original ensemble, and produces projections that are much closer to historical observations (\cref{fig:post_rmse}).

\subsubsection{Second step: Distribution fitting}
\label{sec:final}
\begin{figure}
     \centering
     \begin{subfigure}[t]{5.9cm}
         \centering
         \includegraphics[width=\textwidth]{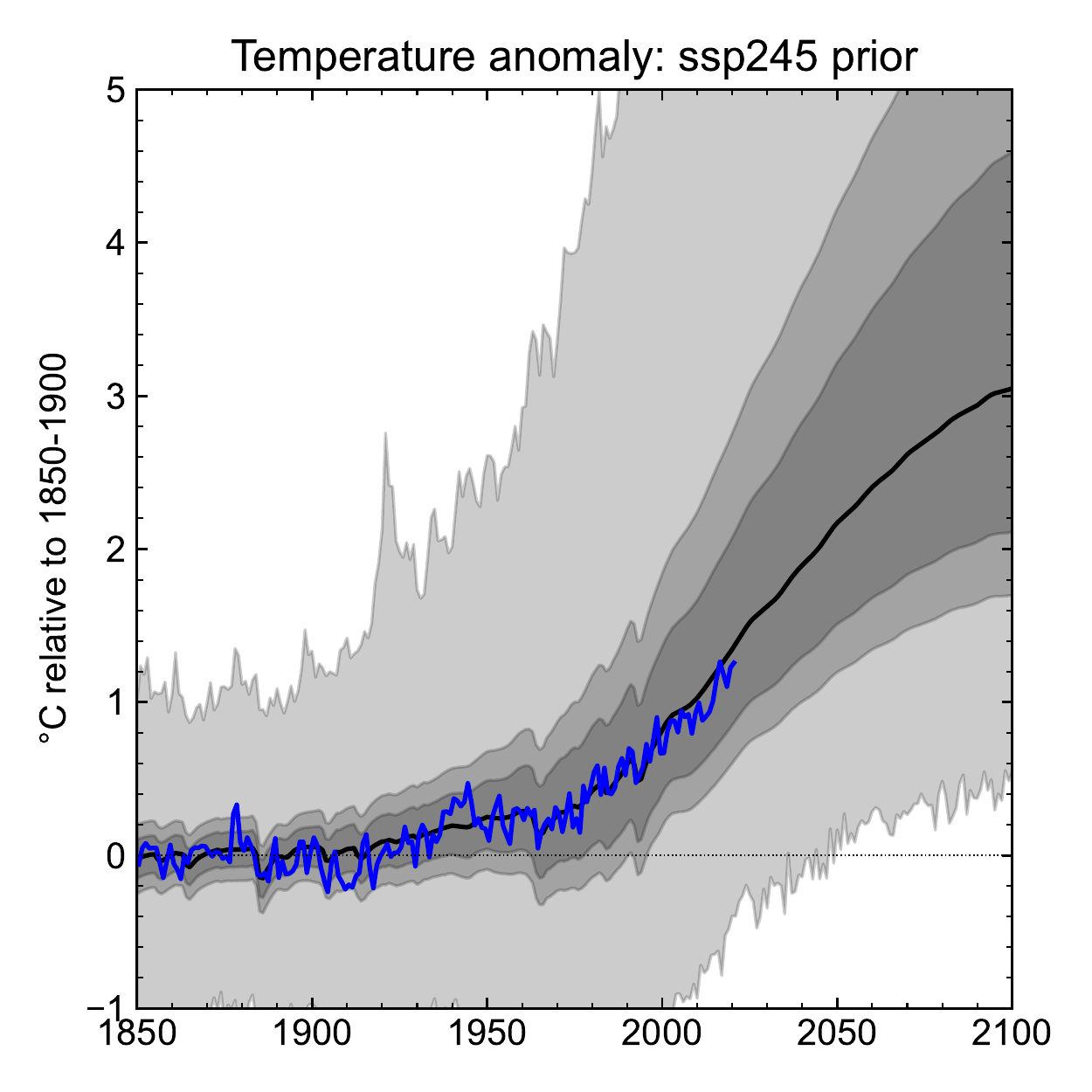}
         \caption{Prior}
         \label{fig:prior}
     \end{subfigure}
     \hfill
     \begin{subfigure}[t]{5.9cm}
         \centering
         \includegraphics[width=\textwidth]{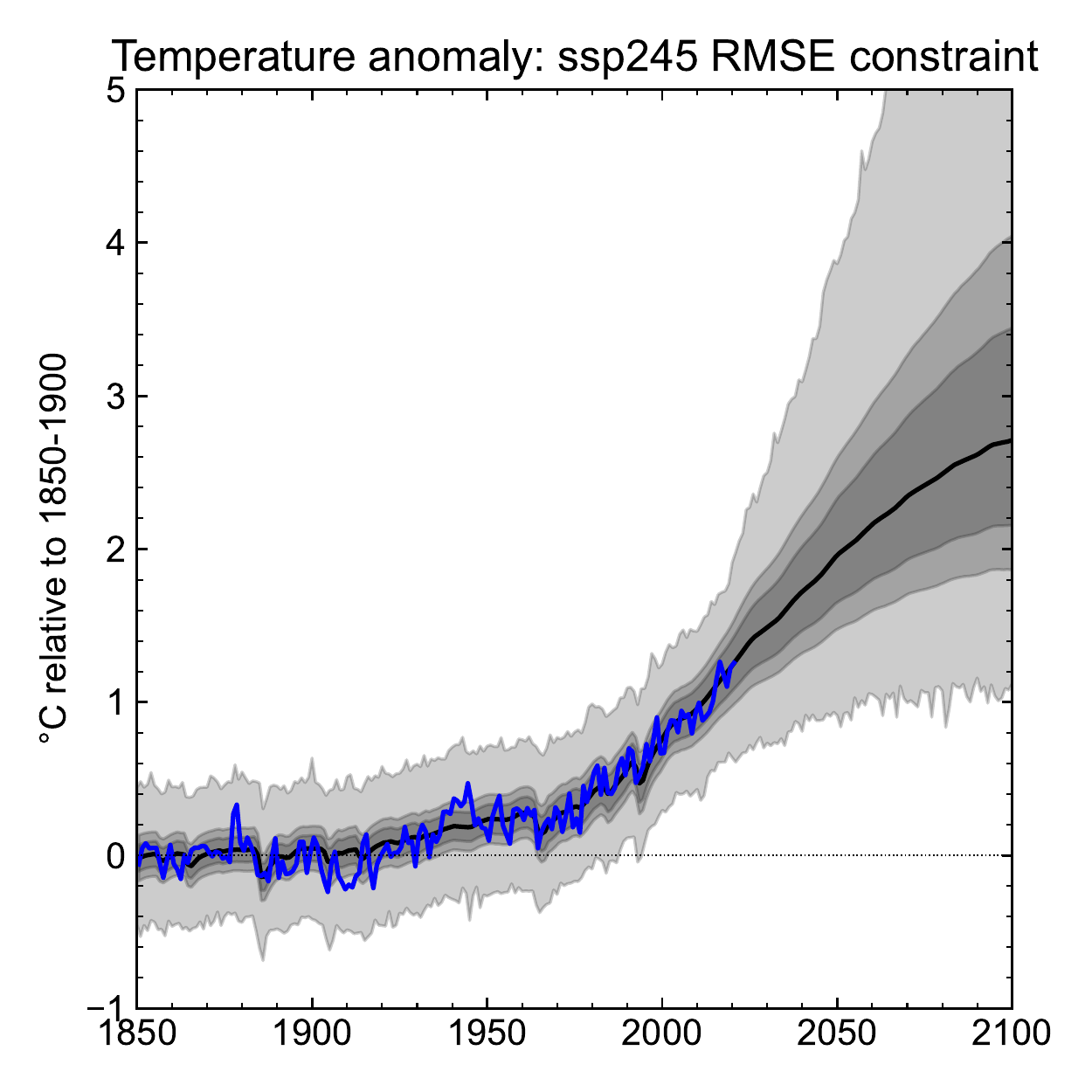}
         \caption{After RMSE constraint}
         \label{fig:post_rmse}
     \end{subfigure}
     \hfill
     \begin{subfigure}[t]{5.9cm}
         \centering
         \includegraphics[width=\textwidth]{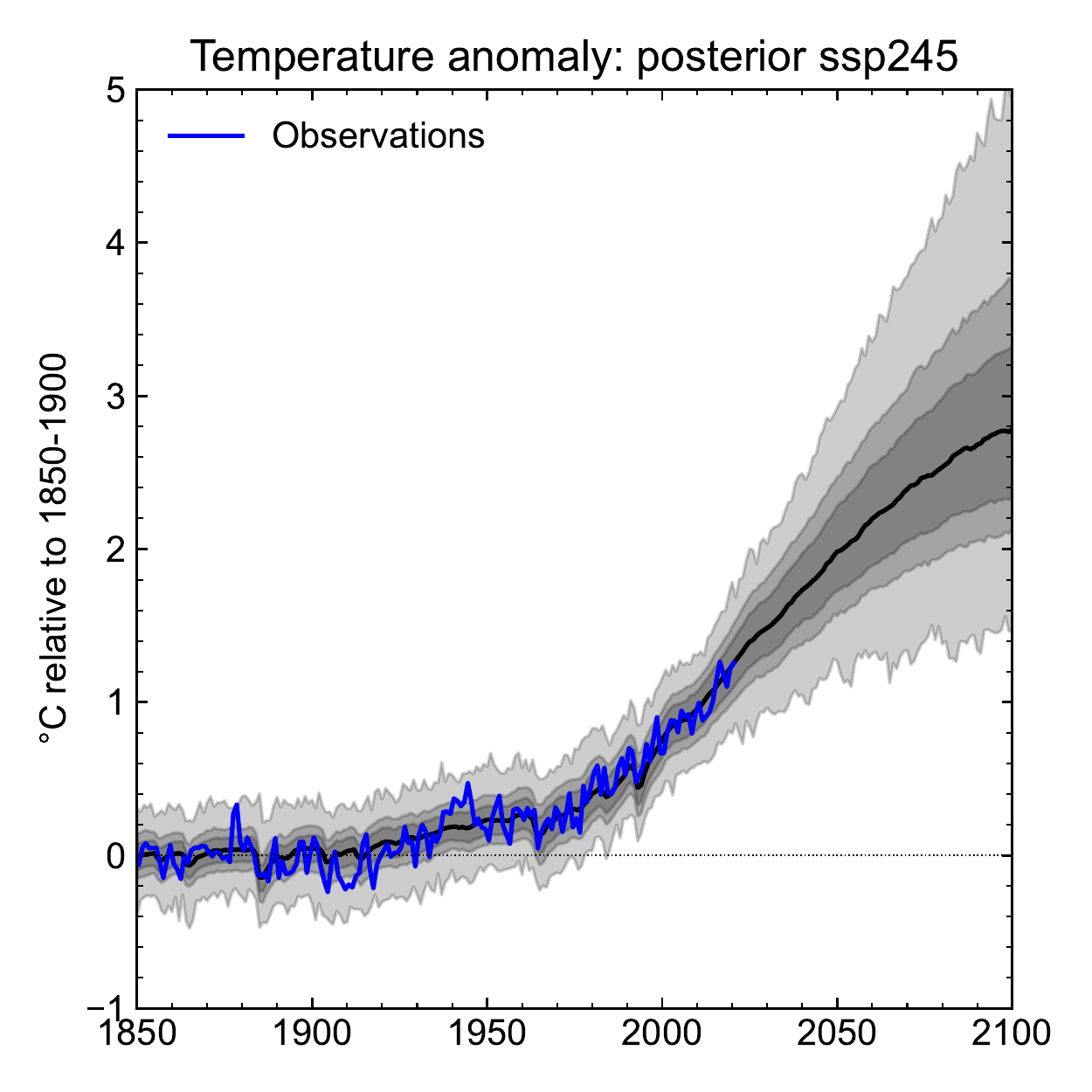}
         \caption{Final reweighed ensemble}
         \label{fig:final}
     \end{subfigure}
        \caption{SSP2-4.5 emissions run in FaIR for the prior, RMSE-only constraint and final, reweighted ensemble. In each subplot, the full minimum to maximum range is in light grey, the 5-95\% range mid-grey, 16-84\% range dark grey and ensemble median black. Historical temperature observations are shown in blue.}
        \label{fig:ssp245}
\end{figure}

The second constraining step takes the ensemble members that pass the RMSE test and attempts to simultaneously fill distributions of observations and climate system metrics that were assessed in the IPCC AR6 WG1 \cite{Smith2021AR6}. The distributions are given in \cref{tab:constraints}. A three-parameter skew-normal distribution is used to fit a distribution to the percentiles given in \cref{tab:constraints} (``Target'' column), which reduces to a Gaussian where the upper and lower ranges are symmetric around the best estimate. Distribution fitting further narrows the range of climate projections (\cref{fig:final}), particularly for future warming.

The size of the final reweighted posterior is user-definable, but is limited by the number of runs passing the RMSE step, which itself depends on the size of the prior. We choose 1001 ensemble members for the final ensemble. This number is sufficiently large to fill the uncertainty space (consistent with constraints) and small enough to be computationally lightweight in the FaIR-DICE coupled system. Typically, ensemble sizes of a few hundred or a few thousand are used in reduced-complexity climate model applications \cite{Nicholls2021}. The final 1001-member constrained, reweighted posterior distribution is, in most cases, close to the target (\cref{tab:constraints} ``Ensemble'' column).

\begin{table*}
\centering
\caption{Target distributions, and percentiles from the reweighted constrained posterior, for climate system assessments used to constrain FaIR. All distributions are from the IPCC AR6 Working Group 1.}
\label{tab:constraints}
\begin{tabular}{l|rrr|rrr}
\multicolumn{1}{l}{} & \multicolumn{3}{|c}{Target} & \multicolumn{3}{|c}{Ensemble} \\
Variable & 5\% & 50\% & 95\% & 5\% & 50\% & 95\% \\
\midrule
ECS [K] & 2.0 & 3.0 & 5.0 & 1.97 & 3.04 & 5.20 \\
TCR [K] & 1.2 & 1.8 & 2.4 & 1.34 & 1.84 & 2.48  \\
Historical warming 1850--1900 to 1995--2014 [K] & 0.67 & 0.85 & 0.98 & 0.71 & 0.85 & 0.99  \\
ERFari 1750 to 2005--2014 [W m$^{-2}$] & $-0.6$ & $-0.3$ & 0.0 & $-0.60$ & $-0.30$ & +0.02  \\
ERFaci 1750 to 2005--2014 [W m$^{-2}$] & $-1.7$ & $-1.0$ & $-0.3$ & $-1.75$ & $-1.04$ & $-0.34$ \\
Total aerosol forcing 1750 to 2005--2014 [W m$^{-2}$] & $-2.0$ & $-1.3$ & $-0.6$ & $-2.06$ & $-1.37$ & $-0.59$ \\
CO$_2$ concentration in 2014 [ppm] & 396.95 & 397.55 & 398.15 & 396.87 & 397.56 & 398.24 \\
Ocean heat content change 1971 to 2018 [ZJ] & 286 & 396 & 506 & 275 & 401 & 514  \\
SSP2-4.5 warming 1995--2014 to 2081--2100 [K] & 1.24 & 1.81 & 2.59 & 1.22 & 1.83 & 2.70  \\
\bottomrule
\end{tabular}
\end{table*}

\begin{figure*}[tbhp]
\centering
\includegraphics[width=\textwidth]{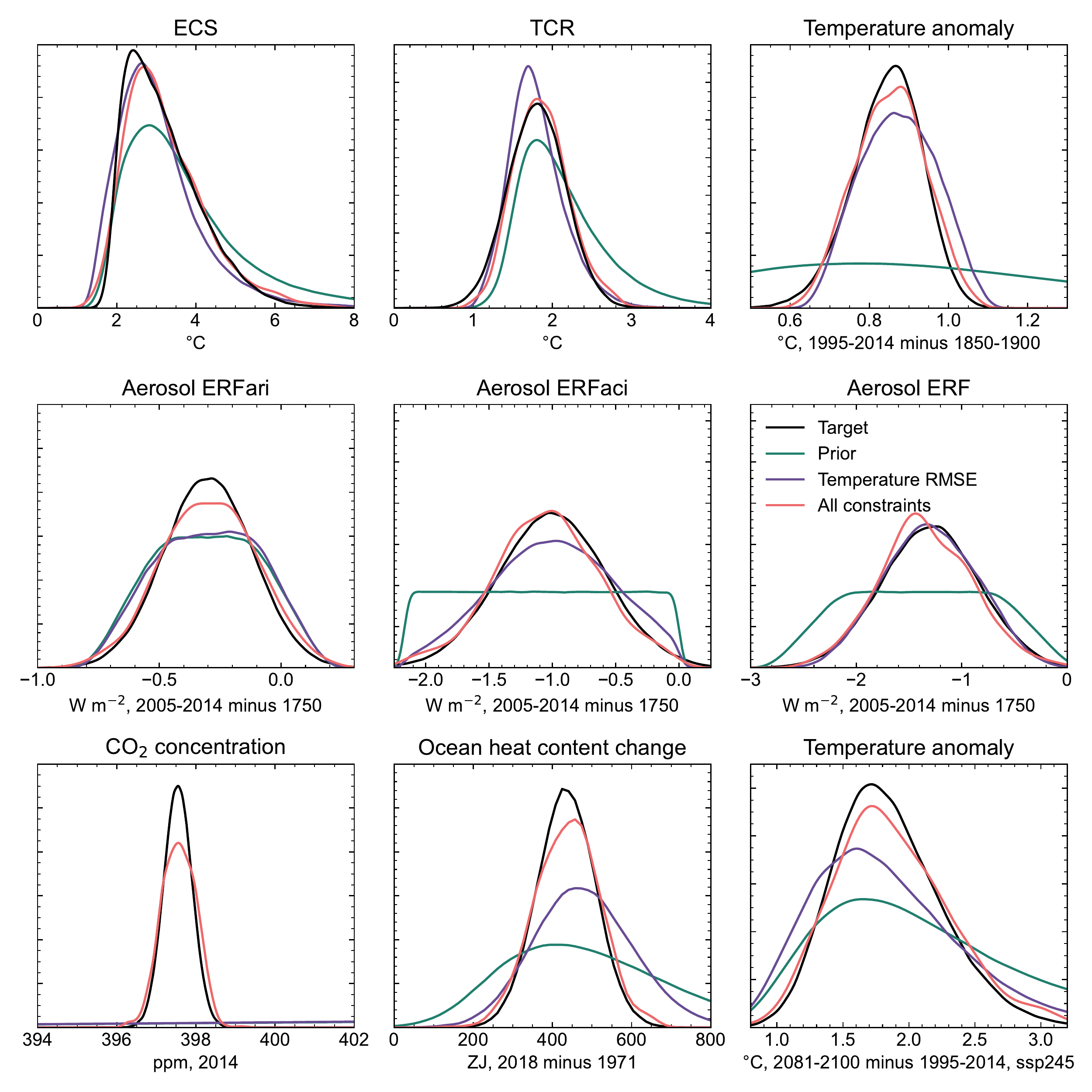}
\caption{Constraining the FaIR posterior to nine IPCC assessments of observable and emergent climate metrics. In each distribution, the green curve shows the prior distribution derived from a 1,500,000 member ensemble in \cref{sec:sampling}. The purple curves shows the distributions after the RMSE filtering step (\cref{sec:rmse}). The red curves show the final reweighted posterior distributions (\cref{sec:final}). The aim is to match the red curves as closely as possible to the target distributions, constructed from \cref{tab:constraints}, in black curves.}
\label{fig:constraints}
\end{figure*}

The distribution matching procedure to produce a constrained ensemble used here is similar and updated from that that was developed as part of the IPCC's Sixth Assessment Report (AR6) WG1, supported by the Reduced Complexity Model Intercomparison Project (RCMIP) \cite{Nicholls2020,Nicholls2021,Forster2021,Smith2021AR6}. This evaluated the ability of simple climate models to reproduce historically observed climate change and expert assessments of emergent climate variables. Models that were fit for purpose were carried forward to provide climate projections from PB-IAM emissions scenarios for the IPCC AR6 WG3 report \cite{Riahi2022,Kikstra2022}. FaIR v1.6.2 was one simple climate model that was deemed to be fit for purpose \cite{Forster2021}. It was not the only one, but of the three that passed the test it is (a) structurally simple and (b) open source (at time of AR6 publication), lending it to be easily used inside the optimization code of a CB-IAM \cite{Haensel2020,Faulwasser2018}.

\section{Historical spin-up and non-CO$_2$ forcing}

In the reduced version of FaIR in DICE, the carbon pools $R_i$, temperatures in each energy balance model layer $T_j$ and non-CO$_2$ effective radiative forcing $F_{\mathrm{ext}}$ need initial values to start from in 2023. Furthermore, $F_{\mathrm{ext}}(t)$ must be provided as an exogenous time series in each ensemble member for 2023--2500.

To do this, we run the 1001-member final calibrated, constrained ensemble from \cref{sec:final} in FaIR, with all forcings (but no internal variability) and uncoupled to DICE, for 1750 to 2500 in 3-year time steps. We run the SSP1-1.9, SSP1-2.6 and SSP2-4.5 emissions scenarios, which provide the non-CO$_2$ forcing time series for the 1.5°C overshoot, well-below 2°C and ``optimal'' scenarios that we run in DICE. These non-CO$_2$ forcing time series are saved out for 2023 to 2500 for each ensemble member and scenario. The values of $T_j$ and $R_i$ are saved out for 2023 for each ensemble member and scenario, and used as initial conditions.

As described in section 2.2 in the main paper the historical all-forcing implementation of FaIR uses a different relationship for calculating ERF from CO$_2$ than the CO$_2$-only DICE-coupled implementation. We also save out the value of $F_{2\times\mathrm{CO}_2}$ used in the computation of future CO$_2$ forcing in FaIR-DICE (\cref{eq:forcing_co2}).

\section{Climate uncertainty is not just ECS}

Consideration of climate uncertainty in CB-IAM analyses may be limited to cases where the eqilibrium climate sensitivity (ECS) is varied but other climate system parameters are held at their best estimate or default values \cite{Dietz2015,IWG2016,Gillingham2018,Yang2021oe}. Notably, there is a strong anti-correlation between ECS and aerosol radiative forcing in historically-consistent climate simulations \cite{Smith2018,Watson-Parris2022}. A high ECS could also be masked by a slow response timescale of the deep ocean or a sea-surface-temperature mediated pattern effect \cite{Dong2020}.

\begin{figure*}[tbhp]
\centering
\includegraphics[width=8.9cm,height=8.9cm]{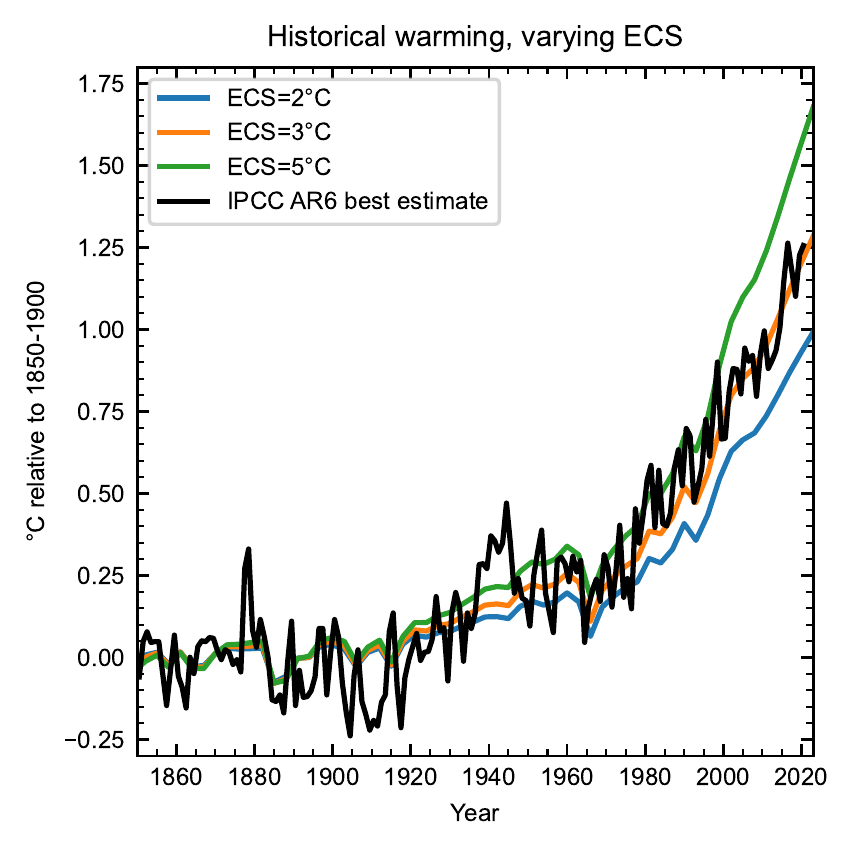}
\caption{Historical climate projected with FaIR with the climate feedback parameter varied to produce ECS=2°C, 3°C and 5°C and all other climate parameters at their ensemble median values from the 1001-member calibrated and constrained ensemble. Observed global mean surface temperature in black.}
\label{fig:ecs_variation}
\end{figure*}

We briefly demonstrate why full climate uncertainty needs to be taken into account, and varying the ECS alone is not sufficient. In \cref{fig:ecs_variation}, we run FaIR over the historical period (1750--2023, at three year time steps) with median parameter values from the 1001-member calibrated, constrained FaIR ensemble, except for the climate feedback parameter $-\kappa_1$ which is inversely proportional to ECS. We choose $-\kappa_1$ to produce ECS=2°C, 3°C and 5°C (5th, 50th and 95th percentiles of the IPCC AR6 distribution). While ECS=3°C (by design) reproduces historical temperatures well, ECS=2°C is too cool, and ECS=5°C is too warm (substantially so). Different parameter combinations would permit historically consistent climates with ECS=2°C (or lower) and ECS=5°C (or higher) but with very different future warming outcomes. Therefore, varying only ECS and not the full spectrum of climate uncertainty can result in historically inconsistent climates which reduces confidence in projections made using climate-economic models.

\section{Comparison of results with DICE-2023 and DICE-2016}
\label{sec:20232016}

We now turn to the beta version of DICE-2023R which incorporates the climate and carbon cycle modules from FaIR v1.0 \cite{Barrage2023}. Alongside the climate module update, DICE-2023R includes more recent updates to the economic and population assumptions over DICE-2016R (as does our version), but the economic part of the IAM is structrually unchanged. For completeness, we also compare our results to DICE-2016R. We compare ``optimal'' scenarios across the three versions of DICE. DICE-2023R and DICE-2016R both produced 2°C and 1.5°C style scenarios but under different definitions to ours and the comparisons are less meaningful. The GAMS code for DICE-2023R and DICE-2016R was obtained from the online appendix to \cite{Barrage2023} (accessed 16 June 2023), and included as part of the software repository that accompanies this paper.

\begin{figure*}[tbhp]
\centering
\includegraphics[width=11.9cm,height=11.9cm]{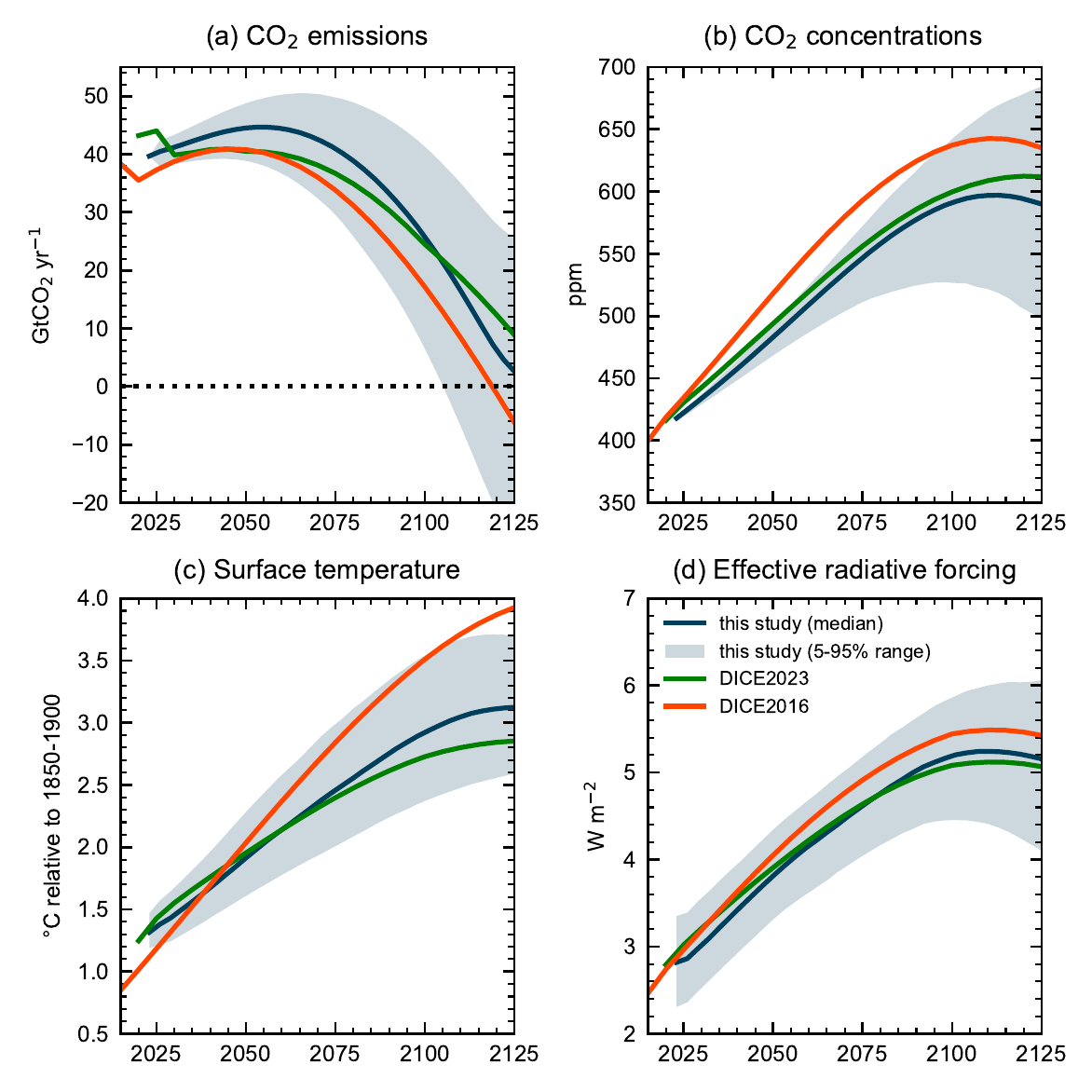}
\caption{Climate projections under the ``optimal'' scenario for the FaIR-DICE implementation used in this study (blue lines ensemble medians; shaded regions 5--95\% ranges) compared to DICE-2023R (green) and DICE-2016R (orange).}
\label{fig:20232016}
\end{figure*}

The comparison of the three models' ``optimal'' scenarios is shown in \cref{fig:20232016}. The improvement in carbon cycle dynamics between DICE-2016R (orange) and DICE-2023R (green) is evident in the trajectory of atmospheric CO$_2$ concentrations (\cref{fig:20232016}b) over the next 100 years. The DICE-2016R concentrations are substantially higher than both our FaIR-DICE (blue) and DICE-2023R despite lower total CO$_2$ emissions (\cref{fig:20232016}a). Nevertheless, the carbon cycle parameterisation in DICE-2023 is more sensitive than in our study given higher CO$_2$ concentrations over the next 100 years despite lower emissions before 2100. As our CO$_2$ concentrations in the present-day are specifically constrained for in the posterior ensemble (main paper fig. 1b; supplement \cref{fig:constraints}) and the carbon cycle models in FaIR-DICE and DICE-2023R are similar, the differences are likely due to a slightly outdated set of carbon cycle parameters used in DICE-2023R.

Temperature (\cref{fig:20232016}c) responds to effective radiative forcing (\cref{fig:20232016}d). The ERF is similar between the three versions, particularly our version compared to DICE-2023R. This leads to fairly similar temperature projections until 2075, at which point DICE-2023R warms slower than our version. DICE-2023R uses a two-layer impulse response model for climate which is similar to our three-layer model, and has the same ECS and TCR as our ensemble median. The differences may arise from the timescales of climate response; DICE-2023R uses FaIR v1.0 default values which are a CMIP5 multi-model mean, whereas we use a constrained probabilistic set drawn from CMIP6 model priors. The temperature response in DICE-2023R is clearly a large improvement over DICE-2016R and well within our uncertainty range.

In summary, DICE-2023R is capturing many of the key behaviours of our version of FaIR-DICE and could be improved further with a re-calibration of the carbon cycle and possibly the climate response. The models differ in other ways including their population and economic assumptions, time step, time horizon, and the fact that DICE-2023R considers non-CO$_2$ greenhouse gases as a separate forcing category. A full comparison of the two models is beyond the scope of this study. 

\section{Comparison to Rennert et al. (2022)}

One additional case we run is attempting to recreate the National Academies of Science, Engineering and Medicine (NASEM) update to the SCC from \citeauthor{Rennert2022} \cite{Rennert2022}. We do this in order to attempt to benchmark our results with a more commonly used discount rate of 2\%. 

\citeauthor{Rennert2022} use a different method to calculate SCC, where 10,000 projections of population (which we adopt, \cref{fig:population}) and emissions of CO$_2$, CH$_4$ and N$_2$O are generated following expert elicitation, and SCC is calculated as the marginal discounted damages from an additional pulse of CO$_2$ into the future. They also use a different integrated assessment model (GIVE) and a different climate damage function to calculate SCC. Climate uncertainty is considered in the overall system and is provided by the IPCC AR6 calibration of FaIR v1.6.2 \cite{Smith2018,Smith2021AR6}.

The discount rate $r$, equivalent to a risk-free interest rate, is given by the Ramsey equation
\begin{equation}
\label{eq:ramsey}
r = \rho + \eta g
\end{equation}
where $\rho$ is the pure rate of time preference, $\eta$ is the marginal utility of consumption and $g$ is the per-capita annual growth in consumption \cite{Ramsey1928}. In order to achieve a near-term discount rate of 2\%, \citeauthor{Rennert2022} calibrated the $\rho$ and $\eta$ parameters to fit future expected evolution in interest rates from the economic literature that matched the desired value of $r$ in the near term. They obtain $\rho = 0.2 \%$, $\eta = 1.24$, implying $g=1.45\%$. 

\begin{figure*}[tbhp]
\centering
\includegraphics[width=17.8cm,height=17.8cm]{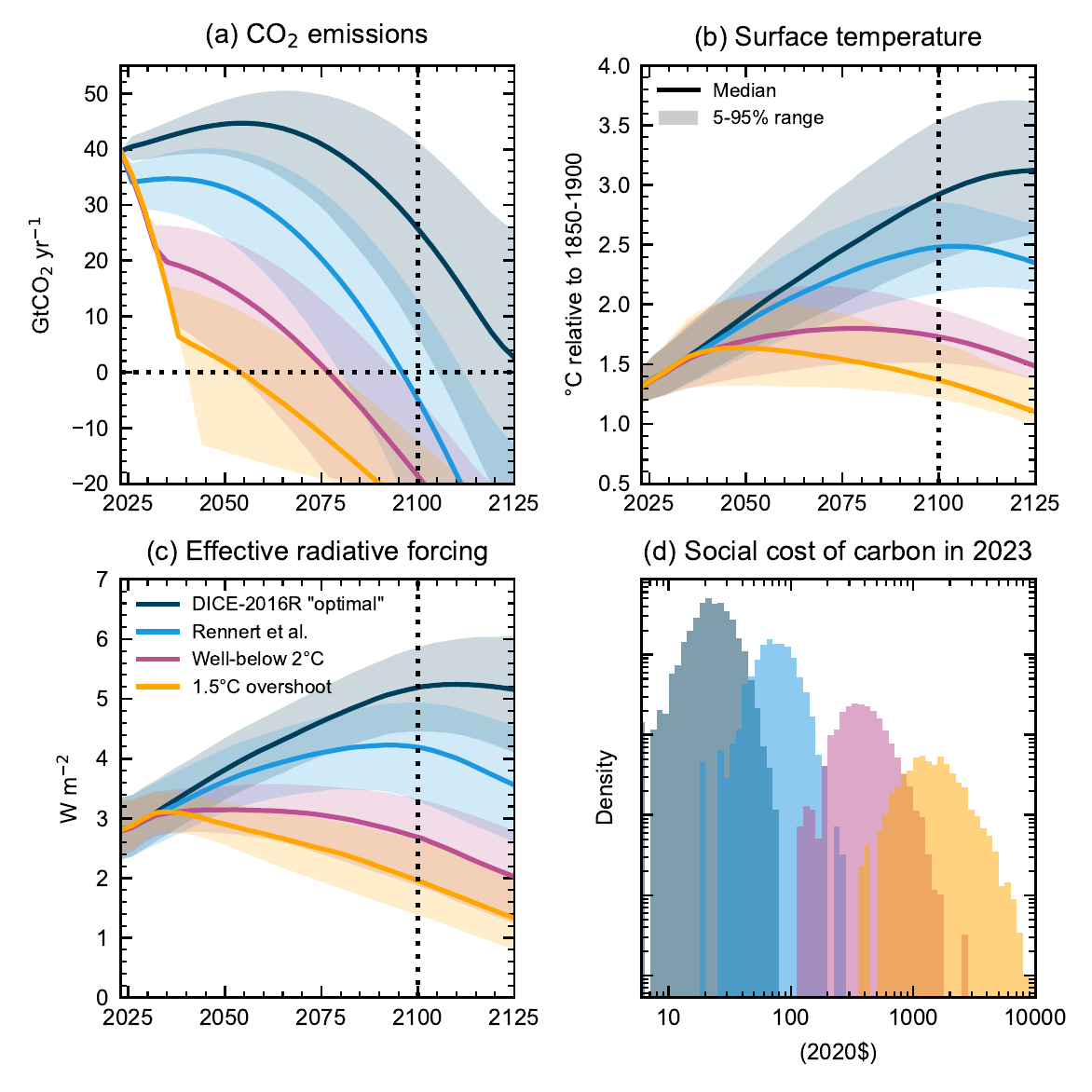}
\caption{Emissions, climate and economic projections for four scenarios using FaIR-DICE, with the additional scenario included that uses the discount parameters from \citeauthor{Rennert2022}. (a) CO$_2$ emissions from energy and industrial processes for  the ``optimal'' (dark blue), \citeauthor{Rennert2022} (light blue), 2°C (pink) and 1.5°C (yellow) scenarios. (b) Temperature projections. (c) Total effective radiative forcing projections. (d) Histogram of year-2023 SCC (in 2020\$) on a log-log scale. In (a-c), light shading shows 5-95\% range and solid lines show ensemble medians.}\label{fig:climate_projections}
\end{figure*}

We re-run our 1001 member ensemble of FaIR-DICE using $\rho = 0.2 \%$, $\eta = 1.24$, retaining SSP2-4.5 as the non-CO$_2$ time series and include this scenario as an additional projection in \cref{fig:climate_projections}. We find this leads us to a median near-term discount rate of 2.5\% and a median SCC of \$84 (t CO$_2$)$^{-1}$ (\cref{tab:results}). Using the NASEM discount parameters leads to an SCC that is less than half of that reported for the 2\% discount case of \$185 (t CO$_2$)$^{-1}$ in \cite{Rennert2022}. \citeauthor{Rennert2022} also show results from a 2.5\% discount rate, where SCC is \$118 (t CO$_2$)$^{-1}$, 40\% higher than our value with the same discount rate.

\begin{table*}
\centering
\caption{Key results from the four scenarios including our implementation of the \citeauthor{Rennert2022} discounting parameters. All correlations are significant at the 1\% level.}
\label{tab:results}
\resizebox{\textwidth}{!}{
\begin{tabular}{lrrrr}
Variable & ``optimal'' & Rennert et al. & Well below 2°C & 1.5°C overshoot \\
\midrule
CO$_2$ emissions 2050 (Gt CO$_2$ yr$^{-1}$) & 45 (39--49) & 33 (24--40) & 15 (2--24) & 2 (\textminus14 to +12) \\
CO$_2$ emissions 2100 (Gt CO$_2$ yr$^{-1}$) & 25 (5--41) & \textminus6 (\textminus21 to +12) & \textminus19 (\textminus23 to \textminus5)   & \textminus23 (\textminus23 to \textminus13) \\
Net zero CO$_2$ year         & 2129 (2105--2152) & 2096 (2080--2112) & 2077 (2053--2094) & 2054 (2040--2079) \\
Social cost of carbon 2023 (2020\$ (t CO$_2$)$^{-1}$) & 26 (15--44) & 84 (49--142) & 439 (237--934) & 1759 (821--4434)\\
Peak warming (°C relative to 1850--1900)         & 3.1 (2.7--3.7) & 2.5 (2.1--2.9) & 1.8 (1.5--2.2) & 1.6 (1.3--2.1) \\
Warming 2100 (°C relative to 1850--1900)          & 2.9 (2.4--3.6) & 2.5 (2.1--2.9) & 1.7 (1.5--2.0) & 1.4 (1.2--1.7) \\
Effective radiative forcing 2100 (W m$^{-2}$)             & 5.2 (4.4--5.9) & 4.2 (3.3--4.9) & 2.7 (1.9--3.3) & 1.9 (1.4--2.6) \\
ECS/SCC correlation coefficient & .51 & .66 & .74 & .74 \\
ECS/2050 CO$_2$ emissions correlation coefficient & \textminus.48 & \textminus.61 & \textminus.72 & \textminus.76 \\
2014 aerosol forcing/SCC correlation coefficient & \textminus.64 & \textminus.63 & \textminus.60 & \textminus.59 \\
2014 aerosol forcing/2050 CO$_2$ emissions correlation coefficient & .61 & .61 & .59 & .56 \\
Near-term discount rate (\%) & 3.1 (3.1--3.2) & 2.5 (2.4--2.6) & 1.4 (1.2--1.6) & 0.6 (0.2--0.8) \\
\bottomrule
\end{tabular}}
\end{table*}

The higher $r$ value that we obtain with the same discounting parameters implies a higher near-term growth rate in our model compared to \cite{Rennert2022}. Indeed, our first to second period per capita growth is 1.84\% per year.

DICE and GIVE are different integrated assessment models, and should not necessarily be expected to give the same results even if the climate system components are the same in both models (which they are not). \citeauthor{Rennert2022} in their Table 1 show that switching from DICE to GIVE, and then updating the damage function from DICE-2016R to theirs increases the SCC by 80\%. Therefore, it is expected that our SCC estimate would be lower than that provided in \cite{Rennert2022} for a 2.5\% discount rate. It should also be noted that achieving an ensemble median 2\% discount rate in DICE would require an iterative approach to adjusting the $\rho$ and $\eta$ parameters, as unlike in GIVE, per-capita growth rates in DICE are dynamic and solved as part of the overall system.

\bibliography{supplement}